\documentclass[a4paper,twocolumn,aps,nofootinbib]{revtex4}
\usepackage[english]{babel}

\usepackage{amssymb,amsthm,mathrsfs,amscd,amsmath}

\usepackage{graphicx,color}
\usepackage{multirow}
\usepackage{array}

\newcommand{\ham}{\hat{\mathcal{H}}} 
\newcommand{\ii}{\mathrm{i}} 
\DeclareMathOperator{\e}{e} 

\begin{document}

\title[Title]{Floquet topological system based on frequency-modulated classical coupled harmonic oscillators}

\author{Grazia Salerno}
\email{grazia.salerno@unitn.it}
\author{Tomoki Ozawa}
\author{Hannah M. Price}
\author{Iacopo Carusotto}
\affiliation{INO-CNR BEC Center and Department of Physics, University of Trento - via Sommarive 14 38123 Povo}

\begin{abstract}
We theoretically propose how to observe topological effects in a generic classical system of coupled harmonic oscillators, such as classical pendula or lumped-element electric circuits, whose oscillation frequency is modulated fast in time. Making use of Floquet theory in the high frequency limit, we identify a regime in which the system is accurately described by a Harper-Hofstadter model where the synthetic magnetic field can be externally tuned via the phase of the frequency-modulation of the different oscillators. We illustrate how the topologically-protected chiral edge states, as well as the Hofstadter butterfly of bulk bands, can be observed in the driven-dissipative steady state under a monochromatic drive. In analogy with the integer quantum Hall effect, we show how the topological Chern numbers of the bands can be extracted from the mean transverse shift of the steady-state oscillation amplitude distribution. Finally we discuss the regime where the analogy with the Harper-Hofstadter model breaks down.

\end{abstract}
\date{\today}
\maketitle

\section{Introduction}

Since topological concepts were first introduced to understand the behaviour of electrons in the quantum Hall effect~\cite{Thouless}, such ideas have become a powerful theme in condensed matter physics, applicable to a wide-range of different systems~\cite{Hasan}. In topological materials, the eigenstates making up an electronic energy band can be characterised by nontrivial topological invariants, which are robust against any perturbations that do not close the energy band gaps. These invariants can be directly related to the existence of remarkable topological edge states localised near the system boundary, which lead, for example, to the robust and precise quantisation of conductance in the quantum Hall effect~\cite{Thouless}. 

While the study of real materials is usually complicated, the essential physics of various topological phenomena can be captured through simple lattice models. Perhaps the most famous of these is the Harper-Hofstadter (HH) model of a charged electron hopping on a 2D tight-binding square lattice in a uniform magnetic field~\cite{Hofstadter}. Not only does this model exhibit the quantum Hall effect thanks to its topological energy bands and unidirectionally-propagating edge states, but it has a rich fractal energy spectrum known as the Hofstadter butterfly. Most importantly, this model has opened up the way for the study of quantum Hall physics beyond traditional solid-state materials, beginning with its realisation in solid-state superlattice devices~\cite{Yu, Dean}, ultracold atomic gases~\cite{MonikaHH,MiyakeHH, syntheticFI, syntheticJQI} and silicon photonics~\cite{Hafezi1,Hafezi2}. 

Recently, topological concepts have entered the realm of classical physics with specially-designed mechanical meta-materials \cite{Vitelli, Kane} and arrays of coupled classical harmonic oscillators such as pendula~\cite{Salerno, Huber}, gyroscopes~\cite{Vitelligyro,Bertoldigyro} and electrical circuits~\cite{Simon, Albert}. For many of these systems,  topological properties are encoded in the normal-mode dynamical matrix, in analogy with the electronic Hamiltonian of a topological lattice model. The demonstration of topological edge states in these classical systems has also ignited the hope that such set-ups could find practical applications, for instance, as loss-free propagating acoustic waveguides~\cite{Huber, Vitelligyro}.  

In neutral systems, such as atoms, photons and classical oscillators, the simulation of topological phenomena has been made possible by important interdisciplinary developments in the creation of so-called synthetic gauge fields~\cite{Dalibard, NathanRev1, NathanRev2, Wang, Raghu}. In the case of the HH model, the effect of the magnetic field on a charged electron is to modify the tight-binding tunnelling amplitudes by complex spatially-dependent phases called the Peierls phases~\cite{Dalibard}. The route to simulating this model with a neutral system is therefore to find an alternative mechanism to generate suitable complex hopping phases. In ultracold gases, this can be, for example, through the control of internal atomic states, by laser-assisted tunneling or lattice shaking~\cite{Dalibard, Struck, Gregor, Kolovsky, Monika, syntheticFI, syntheticJQI}, while in photonics, this has been realised through the insertion of link resonators in ring-resonator arrays~\cite{Hafezi1,Hafezi2} or by twisting waveguides helically along the direction of light propagation to simulate a time-dependent modulation~\cite{Rechtsman}. Finally, in classical systems, analogous effects arise naturally for coupled gyroscopes~\cite{Vitelligyro,Bertoldigyro}, with angular-momentum biasing~\cite{Alu1, Alu2}, or can be hard-wired into the coupling-connections in lattices of pendula~\cite{Huber} or electrical elements~\cite{Simon, Albert}. 

In this paper, we propose a different conceptual approach for how to introduce non-trivial coupling phases, and hence how to simulate topological phenomena, in classical harmonic oscillator lattices through the application of temporally-periodic modulations to system parameters~\footnote{After submission of this manuscript, a preprint \cite{Alu} appeared where a related idea of Floquet topological insulator for sound is discussed.}. In this approach, we build on advances and methods from ultracold gases~\cite{MonikaHH, MiyakeHH, Struck} and photonics~\cite{Rechtsman, Fang, Schmidt, Minkov}, where time-periodic modulations were previously introduced to realise topological lattice models. For classical harmonic oscillators, the first steps in this direction were presented in~\cite{Salerno} for the case of two coupled oscillators; here, we develop these ideas further to propose a modulation scheme that simulates the analogue of the HH model in a large array of coupled oscillators. Even though our discussion is focused on mechanical pendulum systems, the same ideas straightforwardly extend to other types of oscillators such as lumped-element electric circuits of~\cite{Simon}.

As in the gyroscopic lattices of~\cite{Vitelligyro,Bertoldigyro}, our proposal breaks time-reversal symmetry and can be associated with energy bands describing the collective excitation modes of the coupled pendula system, which are characterised by a nontrivial Chern number: the topological invariant of the quantum Hall effect. We go beyond these previous works to show how, not only the edge states, but the hallmarks of the Chern number itself could be seen experimentally in a system of classical harmonic oscillators. While these systems focus on the Haldane model~\cite{Haldane}, our scheme focuses on the HH model, where, thanks to the easy tunability of our synthetic flux, we show that the full Hofstadter butterfly of states could also be observed. Finally, we note that our proposal stands in contrast to the recent experiments on pendula~\cite{Huber} and electrical circuits~\cite{Simon} that did not break time-reversal symmetry, but instead realised two copies of the HH model with opposite synthetic flux: one copy for each of the two ``pseudo-spin" degrees of freedom defined in these systems. In the quantum analogy, this leads to the quantum spin Hall effect in which the edge states are not topologically robust as they are unprotected against perturbations which flip the ``pseudo-spin". 

The paper is organized as follows. In Sec.~\ref{sec:NEM} we introduce the system of coupled classical harmonic oscillators and highlight the limit in which Newton's equations of motion are analogous to the Heisenberg equations for a tight-binding model. In Sec.~\ref{sec:HHm}, we introduce the specific frequency modulation scheme from which we recover the HH model in the high frequency expansion of Floquet theory. The efficiency of our idea is quantitatively assessed in Sec.~\ref{sec:results} where numerical simulations of the full classical equations of motion, are compared to the predictions of the HH model. We find that the oscillation amplitude spectra exhibit wide peaks corresponding to the bulk energy bands; we show that these peaks build up to give the Hofstadter butterfly as a function of the synthetic magnetic flux. We excite topologically-protected edge states showing that these are chiral and immune from back-scattering. We then discuss a classical counterpart of the integer quantum Hall effect from which we can extract the Chern number. In Sec.~\ref{sec:instability}, we investigate the dynamical stability of the system and the role of the counter-rotating-wave terms, together with the consequent limitations of the mapping onto the HH model. We comment in Sec.~\ref{sec:experimental} on possible experimental implementations of our proposal using arrays of either mechanical pendula or electric circuits, before drawing conclusions in Sec.~\ref{sec:conclusions}.

\section{Newton's equations of motion}
\label{sec:NEM}
\begin{figure}[t]
\centering
\includegraphics[width=0.4\textwidth]{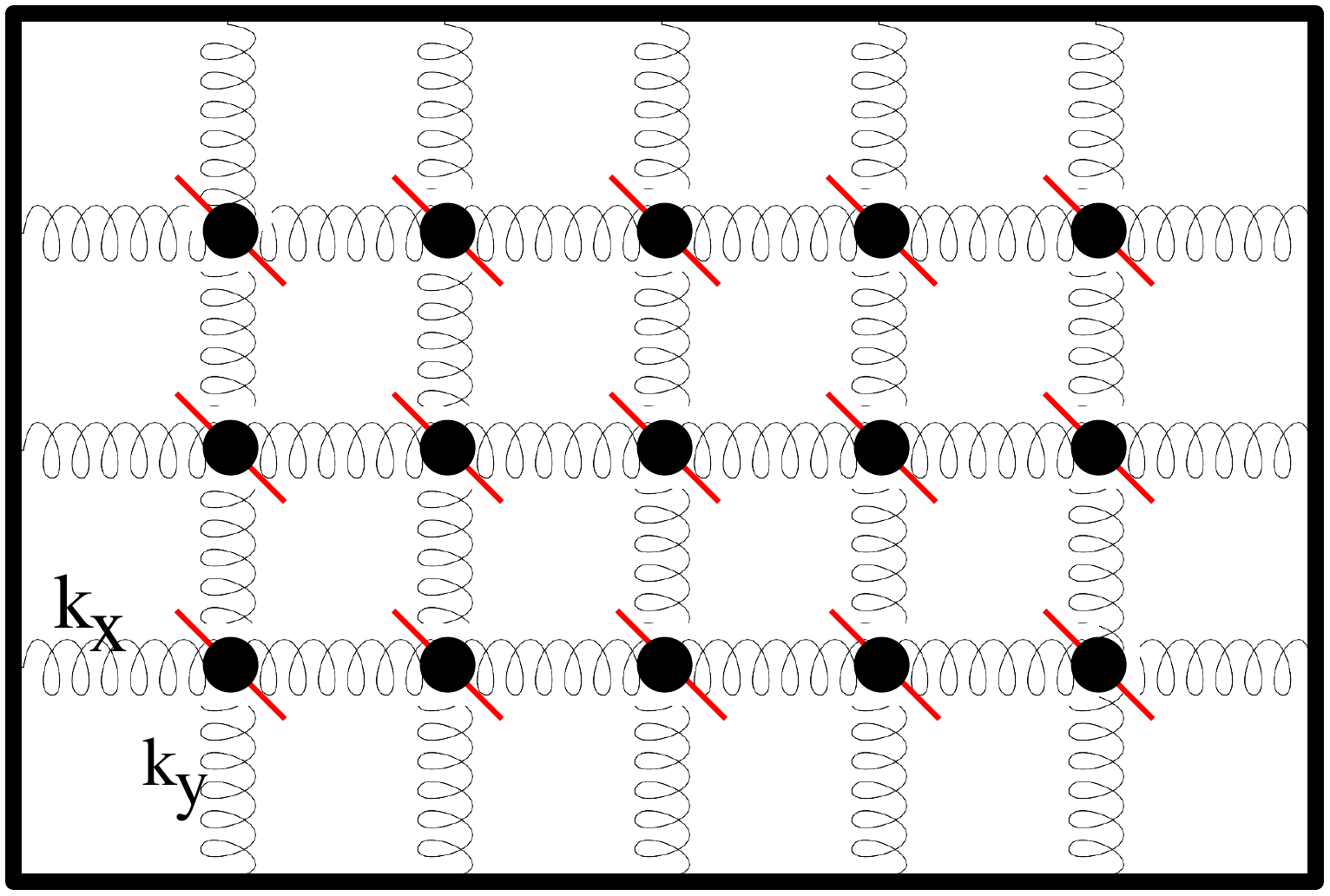}
\caption{View from above of a square lattice of pendula constrained to oscillate along the plane indicated by the thick red lines. The coupling between neighbouring pendula is provided by elastic springs with spring constants $k_x$ and $k_y$. Pendula on the edges are coupled to a rigid wall.}
\label{fig:system}
\end{figure}

We consider a square lattice of $N_x\times N_y$ sites. Each lattice site, labelled by two indexes $(i,j)$, hosts a pendulum of mass $m_{i,j}$ and frequency $\omega_{i,j}(t)$, whose oscillation plane is rigidly fixed, for example, along the unit-cell diagonal, as shown in Fig.~\ref{fig:system}. We assume that the natural frequencies of the pendula are periodically modulated in time around their mean value $\tilde{\omega}_{i,j}\equiv\langle\omega_{i,j}(t)\rangle$. The displacement of each pendulum from its equilibrium point along the fixed direction is denoted by $z_{i,j}$. 

All the pendula are coupled to their nearest neighbours, while the pendula on the edges of the lattice are coupled to a fixed wall. The coupling is provided, for example, by springs of rest length equal to the lattice spacing and of constant $k_x$ and $k_y$ along the two $x$ and $y$ directions respectively. In the small oscillation regime the elongation of each spring is proportional to the difference of the spatial displacement of the pendula that it connects $z_{i,j}-z_{i',j'}$ and, in absence of any pre-tension, the force is determined by the standard Hooke's law.

Including loss mechanisms with a friction coefficient $\xi$, Newton's equations of motion for a pendulum driven by the external force $\mathcal{F}_{i,j}^\text{ex}(t)$ read:
\begin{align}
&m_{i,j} \dot{z}_{i,j} = p_{i,j} \label{motion1}\\
\begin{split}
&\dot{p}_{i,j}=-m_{i,j} \,\omega_{i,j}^2(t) z_{i,j} -2 \xi \dot{z}_{i,j} -\mathcal{F}_{i,j}^\text{ex}(t)\\&\qquad +\sum_{\pm1}k_x \left(z_{i\pm1,j}-z_{i,j} \right) +\sum_{\pm1}k_y \left(z_{i,j\pm1}-z_{i,j} \right).\label{motion2}
\end{split}
\end{align}

The goal of this section is to show that in a suitable regime Newton's equations of motion \eqref{motion1}-\eqref{motion2} can be rewritten in the same form as the Heisenberg equation for a tight-binding Hamiltonian, where the bosonic annihilation and creation operators are replaced by $\mathbb{C}$-numbers. Starting from the displacements $z_{i,j}$ and the corresponding momenta $p_{i,j}$, classical complex variables corresponding to the annihilation operators of quantum harmonic oscillator are defined as:
\begin{equation}
\alpha_{i,j}=\sqrt{\frac{m_{i,j}\,\tilde{\omega}_{i,j}}{2}}\, z_{i,j} + \ii \frac{p_{i,j}}{\sqrt{2m_{i,j}\,\tilde{\omega}_{i,j}}}.
\label{transformation}
\end{equation}
It is straightforward to see that the square modulus $|\alpha_{i,j}|^2$ is proportional to the oscillation energy of the $(i,j)$-th pendulum. 

Assuming for simplicity that the product $m_{i,j}\tilde{\omega}_{i,j}\equiv\mu$ is constant for every site $i,j$ and defining $\Omega_x \equiv k_x/(2\mu)$ and $\Omega_y \equiv k_y/(2\mu)$, we can combine Eq.~\eqref{motion1} and Eq.~\eqref{motion2} into equations for the complex amplitudes $\alpha_{i,j}$ of the form:
\begin{equation}
\begin{split}
\dot{\alpha}_{i,j}= &-\ii \left[ \frac{\tilde{\omega}_{i,j}}{2}+ \frac{\omega_{i,j}^2(t)}{2\tilde{\omega}_{i,j}}-\ii \gamma_{i,j} +2\Omega_x+2\Omega_y\right] \alpha_{i,j} \\ &-\ii \left[ \frac{\tilde{\omega}_{i,j}}{2}- \frac{\omega_{i,j}^2(t)}{2\tilde{\omega}_{i,j}}+\ii \gamma_{i,j} +2\Omega_x+2\Omega_y\right] \alpha^*_{i,j}\\ &
+\ii \sum_{\pm1}\Omega_x\left(\alpha_{i\pm1,j}+\alpha^*_{i\pm1, j}\right) \\&
+ \ii \sum_{\pm1}\Omega_y\left(\alpha_{i,j\pm1}+\alpha^*_{i,j\pm1}\right)+ \ii f_{i,j}^\text{ex}(t),
\end{split}
\label{alphadrivendot}
\end{equation}
where the damping rate $\gamma_{i,j}\equiv \xi/m_{i,j}$. The external driving force is taken to be monochromatic at frequency $\omega_\text{ex}$ with amplitude proportional to $f_{i,j}$,
\begin{equation}
f_{i,j}^\text{ex}(t)\equiv \mathcal{F}_{i,j}^\text{ex}(t)/\sqrt{2\mu}=2 f_{i,j}\cos(\omega_\text{ex} t). 
\end{equation}

Equations~\eqref{alphadrivendot} can be simplified under the assumption that the natural frequencies $\tilde{\omega}_{i,j}$ of the pendula are much larger than all other frequencies in the problem and that the driving frequency $\omega_\text{ex}$ is comparable to the $\tilde{\omega}_{i,j}$'s. In this case, the fastest time dependence of $\alpha_{i,j}$ is proportional to $\e^{-\ii\tilde{\omega}_{i,j}t}$, while that of the conjugate variables $\alpha_{i,j}^*(t)$ is proportional to $\e^{\ii\tilde{\omega}_{i,j}t}$. It can be seen that in the limit $\tilde{\omega}_{i,j}\rightarrow \infty$ the contribution of the $\alpha^*$'s to the motion equation for $\alpha$ is negligible as the fast oscillations quickly average to zero~\footnote{This is most easily understood by moving to a rotating frame at $\tilde{\omega}_{i,j}$, in which the $\alpha$ terms only have a (relatively) slow time-evolution while the $\alpha^*$ terms oscillate at $-2\tilde{\omega}_{i,j}$ and therefore quickly average to zero.}. As a result, it is legitimate to neglect in Eq.~\eqref{alphadrivendot} the terms involving the complex conjugate variables $\alpha^*$, an approximation that in Quantum Optics is well-known under the name of the Rotating Wave Approximation (RWA). 

Applying the RWA, we interpret Eq.~\eqref{alphadrivendot} without driving and dissipation, as the classical analogue of the Heisenberg equation of motion obtained from the following Hamiltonian:
\begin{equation}
\begin{split}
\mathcal{H}(t)=& \sum_{i,j} \left( \frac{\tilde{\omega}_{i,j}}{2}+ \frac{\omega_{i,j}^2(t)}{2\tilde{\omega}_{i,j}} +2\Omega_x+2\Omega_y\right) \alpha^*_{i,j}\alpha_{i,j} \\&- \sum_{i,j} \left( \Omega_x \alpha^*_{i,j} \alpha_{i+1,j} + \Omega_y \alpha^*_{i,j} \alpha_{i,j+1} +\text{c.c.}\right).
\end{split}
\label{tbh}
\end{equation}
This is a bosonic tight-binding Hamiltonian, where the terms on the first line account for the on-site energy and those on the second line describe the hopping. 
In the Hamiltonian, the RWA consists of neglecting all non-energy-conserving terms proportional to $\alpha_i \alpha_j$ and  $\alpha^*_i \alpha^*_j$ and only keep the energy-conserving ones $\alpha^*_i \alpha_j$. 

\section{Effective Harper-Hofstadter model}
\label{sec:HHm}

In the previous Section, we have shown how Newton's equations of motion~\eqref{alphadrivendot} can be rewritten in the RWA in the same form as the Heisenberg equation of a tight-binding model. In this Section we proceed to identify a suitable modulation scheme for the $\omega_{i,j}(t)$ that reduces to an effective HH model in the Floquet picture.

\subsection{Temporal and spatial modulation of the natural frequencies}

Inspired by the realization of the HH model with ultra-cold atoms in optical lattices \cite{MonikaHH, MiyakeHH} and by related proposals in the photonic context~\cite{modulatedCircuit, Fang}, we take the natural frequency of the pendula, \textit{i.e.} a part of the single-particle on-site energy in the tight-binding analogy, to be temporally and spatially modulated according to:
\begin{equation}
\omega_{i,j}^2(t)= \tilde{\omega}_{i,j}^2 \left(1+ 2 \frac{V_{i,j}(t)}{\tilde{\omega}_{i,j}}\right)
\label{barefrequency}
\end{equation}
with a spatial dependence
\begin{equation}
\tilde{\omega}_{i,j}=\omega_0-2\Omega_x -2 \Omega_y + w S(i).
\label{staticmod}
\end{equation} 

\begin{figure}[t]
\centering
\includegraphics[width=0.45\textwidth]{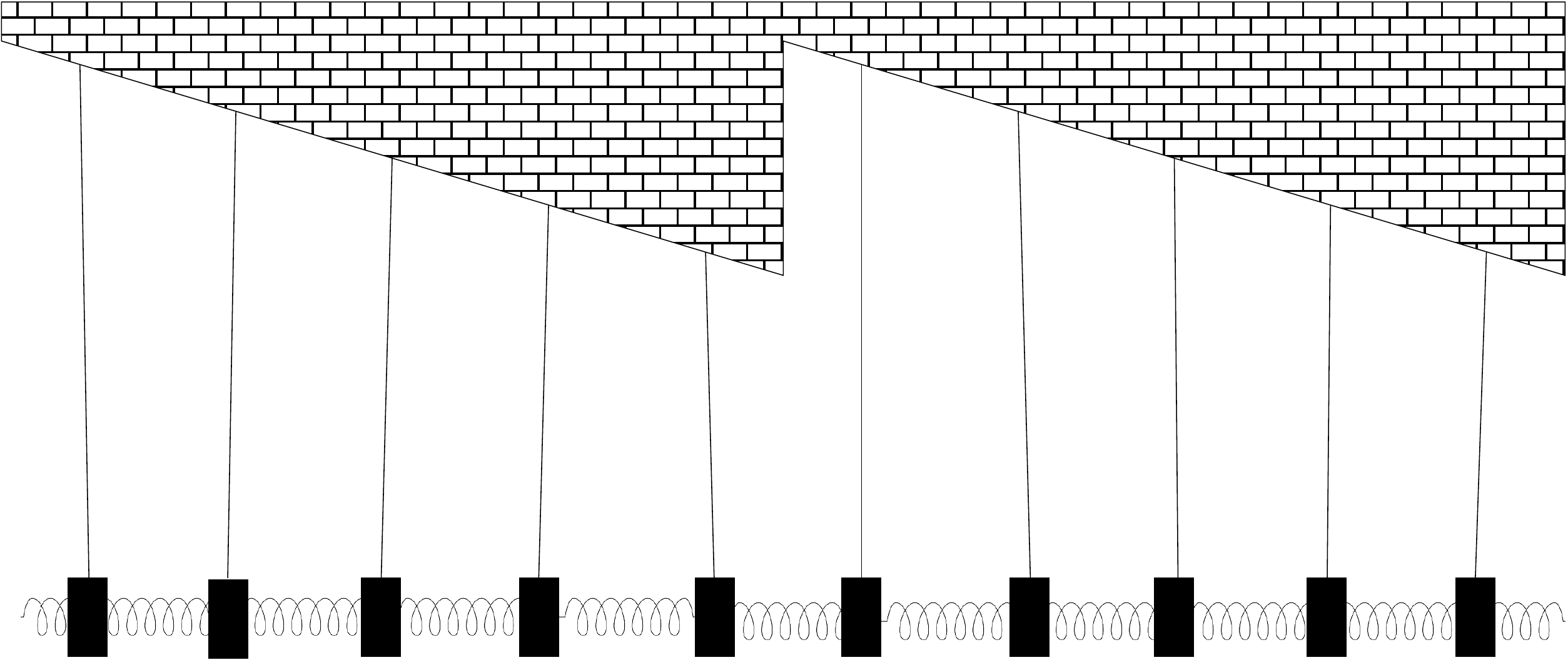}
\caption{Section of the system along the $x$ direction, sketching the spatial modulation of the natural frequencies of the pendula with period $s=5$.}
\label{fig:system_x}
\end{figure}

The temporal modulation $V_{i,j}(t)$ is taken as periodic with frequency $w$, and the static spatial modulation:
\begin{equation}
S(i)\equiv \text{mod}(i-1,s)
\label{S}
\end{equation}
is a saw-tooth function along the $x$-direction, of period $s$, as schematically shown in Fig.~\ref{fig:system_x}. 
While hopping along $y$ is unaffected by both the spatial and temporal modulations, bare hopping along $x$ would be strongly suppressed by the spatial modulation Eq.~\eqref{staticmod}. Hopping in this direction is however restored by the temporal modulation: when the frequency of the modulation is chosen to be on resonance with the detuning of neighbouring sites, a uniform coupling is restored over the whole lattice~\cite{Goldman}. Given the specific sawtooth form of Eq.~\eqref{S} chosen for $S(i)$, a bi-harmonic temporal modulation is needed:
\begin{equation}
\begin{split}
V_{i,j}(t)=V\,\Bigg[&\cos(w t +\phi_{i,j})+ \\ & \; (s-1)\cos\bigg(w(s-1)t-\phi_{i,j}\bigg)\Bigg]
\end{split}
\label{timemodulation}
\end{equation}
where the amplitude $V$ is position independent. The first component at frequency $w$ restores a uniform hopping between the pairs of pendula with natural frequency difference $w$, while the second component at frequency $w(s-1)$ addresses the pairs with a frequency difference $w(s-1)$. As we shall see in the next Subsection, the position-dependence of the phase $\phi_{i,j}$ offers us a way to control the phase of the hopping along $x$~\cite{Monika}. In order to obtain the desired HH model, a possible choice is the following:
\begin{equation}
\phi_{i,j}=2\pi \theta (i+j).
\label{phase}
\end{equation}

\subsection{Analytical derivation of the Harper-Hofstadter effective equation of motion}
With the modulations introduced in the previous paragraph, we now show how the equations of motion in Eq.~\eqref{alphadrivendot} take the form of the Heisenberg equations of motion for the HH model by applying the Floquet formalism of time-periodic systems at high-frequency \cite{Bukov}.

Substituting the chosen forms of the temporal and static modulation Eq.~\eqref{barefrequency} and Eq.~\eqref{staticmod} into Eq.~\eqref{alphadrivendot} and performing the RWA, we straightforwardly obtain:
\begin{equation}
\begin{split}
\dot{\alpha}_{i,j}&=-\ii \left[\omega_0 + w S(i) +V_{i,j}(t) -\ii \gamma_{i,j}\right] \alpha_{i,j}\\& +\ii \sum_{\pm 1} \left[\Omega_x \alpha_{i\pm1, j}+ \Omega_y \alpha_{i,j\pm1}\right] +  \ii f_{i,j}^\text{ex}(t).
\end{split}
\label{alphaRWAdot}
\end{equation}
In the following we focus on the case where a single $(i_p,j_p)$ site is driven. We also assume that the fastest frequency of the temporal modulation $sw$ is fast compared to all other frequencies except the bare natural oscillation frequency:
\begin{equation}
\omega_0 \gg s w \gg \Omega_{x,y}.
\label{inequality}
\end{equation}
The first inequality is required by the RWA, while the latter justifies our calculating the effective dynamics after a time-average over the period $T=2\pi/w$ of the temporal modulation. In this regime, a simple expression for the time-evolution of the whole system at stroboscopic times $t_n = nT$ can be obtained using Floquet theory~\cite{Holthaus}. To this purpose we introduce new variables:
\begin{equation}
\begin{split}
\beta_{i,j}(t)\equiv \,&\alpha_{i,j}(t) \e^{\ii wt \left(S(i) - S(i_p)\right) } \\ &\e^{\ii \omega_\text{ex} t} \exp\left(\ii \int_0^t V_{i,j}(t') \mathrm{d}t'\right).
\label{beta}
\end{split}
\end{equation}
We then look for the steady-state solution, where $\beta_{i,j}$ oscillates with the same frequency as the external driving force, which is tuned in the vicinity of the natural frequency of the driven pendulum $(i_p, j_p)$. An effective equation of motion can be obtained for the $\beta_{i,j}$ variables by time-averaging Eq.~\eqref{alphaRWAdot} over one period of the fast temporal modulation, that is by applying the Magnus expansion to the lowest order \cite{GoldmanX,Bukov,Citro}. 

After some calculations, time-averaged equations for $\beta_{i,j}$ are found. The full forms of the coefficients of the effective equations are expressed as sums of Bessel functions and can be found in the Appendix \ref{app:fulleff}. A much simpler expression for the effective equations of motion is obtained by considering only the terms that give the largest contribution in these sums:
\begin{equation}
\begin{split}
\dot{\beta}_{i,j}&=-\ii \left(\omega_0 +w S(i_p) -\omega_\text{ex}\right) \beta_{i,j} \\&+ \ii \sum_{\pm1} \Omega_x \e^{\mp\ii (\varphi_{i\pm1,j}-\frac{\pi}{2})} \mathcal{J}_{\pm1}\left(I_0\right)\mathcal{J}_{0}\left(I_0\right)\beta_{i\pm 1,j} \\ &+\ii \sum_{\pm1} \Omega_y\mathcal{J}_0\left(I_0\right)^2 \beta_{i,j\pm 1}  \\&+\ii f_{i_p,j_p}^\text{ex}\mathcal{J}_0\left(V/w\right)^2  - \gamma_{i,j} \beta_{i,j}
\label{betadot}
\end{split}
\end{equation}
where, with the definition of the modulation phase in Eq.~\eqref{phase}, we have introduced the following quantities:
\begin{align}
&\varphi_{i\pm1,j}\equiv 2\pi\theta (i+j) \pm \pi\theta \label{fase}\\
&I_0\equiv \frac{V}{w} \sqrt{2-2 \cos(2\pi\theta)}. \label{argument}
\end{align}
As detailed in Appendix \ref{app:fulleff}, the condition of validity for Eq.~\eqref{betadot} is that the next order of the Bessel functions can be neglected, 
\begin{equation}
\mathcal{J}_{(s-1)}(I_0) \ll \mathcal{J}_1(I_0).
\end{equation}
From this expression, it is apparent that a very large period of the spatial modulation $s$ would be optimal, with a linear ramp in the bare frequencies of the harmonic oscillators in Eq.~\eqref{staticmod}. While this configuration is possible for ultracold atomic systems~\cite{MonikaHH,MiyakeHH}, it is not possible for these classical harmonic oscillators as the condition $\omega_0 \gg sw$ for the RWA does not allow us to use an arbitrarily large value of $s$. More details of these issues can be found in Sec.~\ref{sec:experimental}. In the following, we shall see that a good compromise can be found already for $s$ as small as $5$.

In Eq.~\eqref{betadot}, the bare coupling frequencies $\Omega_x$ and $\Omega_y$ are renormalised by the Bessel functions, and the effective couplings along the two directions are different. In order to recover the usual HH model with equal hopping in the two directions, it is useful to start from suitably chosen $\Omega_{x,y}$ that will give the same effective $J$ in the two directions:
\begin{equation}
\Omega_x= \frac{J}{\mathcal{J}_{0}\left(I_0\right)\mathcal{J}_{1}\left(I_0\right)}, \qquad
\Omega_y= \frac{J}{{\mathcal{J}_0\left(I_0\right)}^2}.
\label{renormalisedOmega}
\end{equation}
One must of course be careful in choosing the amplitude of the modulation $V$, such that the frequencies in Eq.~\eqref{renormalisedOmega} satisfy the inequality of Eq.~\eqref{inequality}. 

As with the hopping terms, so is the amplitude of the driving term in Eq.~\eqref{betadot} renormalised by the Bessel functions. We choose the bare driving force on the pumped site as:
\begin{equation}
f_{i_p,j_p}^\text{ex} = 2 f / \mathcal{J}_0 \left( V/w \right)^2
\end{equation}
in order to keep the effective driving intensity $f$ constant as we vary $V$ or $w$. 

Without the last two terms describing driving and dissipation, Eq.~\eqref{betadot} is the classical version of the Heisenberg equations derived from the HH Hamiltonian:
\begin{equation}
\begin{split}
\ham=&\sum_{i,j}\Biggl[-\Delta \omega \hat{\beta}_{i,j}^\dagger \hat{\beta}_{i,j} -J \Bigl(\hat{\beta}_{i,j}^\dagger \hat{\beta}_{i,j+1} +\\&\quad \e^{-\ii 2\pi\theta(i+j)} \e^{-\ii (\pi\theta-\pi/2)} \hat{\beta}_{i,j}^\dagger \hat{\beta}_{i+1,j} + \text{h.c.}\Bigr)\Biggr]
\end{split}
\label{HHH}
\end{equation}
where $\Delta \omega \equiv \omega_\text{ex} -\omega_0 -w S(i_p)$ is the energy at which the HH model is coherently probed.
The non-trivial hopping phase in Eq.\eqref{HHH} is determined by the modulation phase in Eq.~\eqref{phase} and controlled by the parameter $\theta$. As usual, the magnetic flux enclosed within a plaquette is calculated by summing the phases accumulated on each link when hopping around a plaquette of the lattice. With the definition in Eq.~\eqref{phase}, the sum of the phases gained by the complex hopping elements in Eq.~\eqref{betadot}, is uniform for the whole lattice and equal to:
\begin{equation}
\sum_{\square}\phi=\frac{1}{2} \left(-\phi_{i+1,j}-\phi_{i,j}+\phi_{i,j+1}+\phi_{i+1,j+1}\right)=2\pi\theta.
\label{plaquette}
\end{equation}
As a result, our model of coupled pendula is a classical simulator of the HH model and allows us to generate the Hofstadter butterfly by simply tuning the phase of the temporal modulations according to Eq.~\eqref{phase}. The tunability of this artificial magnetic field is a main advantage of our scheme compared to other systems, where the magnetic flux is fixed \cite{Hafezi1,Huber}. We note that any other choice of the temporal modulation phase $\phi_{i,j}$ leading to the same flux per plaquette would give an equivalent HH model that only differs by a gauge transformation.  

Before proceeding, it is important to note that the analysis we have done so far is strictly valid only in the deep RWA limit $\omega_0 \rightarrow \infty$ where the counter-rotating-wave terms $\alpha_{i,j}^*$ in Eq.~\eqref{alphadrivendot} can be safely neglected and effective equations such as Eq.~\eqref{betadot} can be derived. 
For lower values of $\omega_0$, the first effect beyond the RWA is a global shift in the detuning frequency: $\Delta\omega \equiv \omega_\text{ex}-\omega_0 -w S(i_p) \rightarrow \omega_\text{ex}-\bar{\omega}$, with:
\begin{equation}
\bar{\omega}=\omega_0+w S(i_p)-\frac{V^2}{4\,\omega_0}(2-2s+s^2)- 2\frac{(\Omega_x+\Omega_y)^2}{\omega_0}.
\label{shift}
\end{equation}
This shift is predicted by including the counter-rotating-wave part of the $\alpha_{i,j}^*$ in Eq.~\eqref{alphadrivendot}, that in the frame rotating at $\omega_0$ corresponds to a non-rotating term and so includes the leading order of the counter-rotating-wave effects. As discussed in~\cite{Salerno}, this has the form of a classical analogue of the Bloch-Siegert shift. For even smaller $\omega_0$, the counter-rotating-wave terms can also lead to dynamical instabilities and will be discussed in Sec.~\ref{sec:instability}.

\section{Numerical simulations and discussions} 
\label{sec:results}

In order to assess the validity of the effective HH model, obtained via the Floquet approach, we have performed numerical simulations of the full equations of motion~\eqref{alphadrivendot} including the counter-rotating-wave terms as well as the full spatial and temporal modulations of the oscillation frequencies. We have quantitatively compared these predictions to those from the Heisenberg equation of motion for the HH model in Eq.~\eqref{betadot}.

We have numerically integrated the set of $N=N_x\times N_y$ time-dependent equations~\eqref{alphadrivendot} for a square lattice of $N_x\times N_y$ sites with a standard fourth-order Runge-Kutta until a steady state is achieved at long evolution times $t \gg 1/\min(\gamma_{i,j})\equiv 1/\gamma$. To match with the Floquet picture, we have performed a stroboscopic sampling at times $t_n=2\pi n/w$. At these times $t_n$, under a monochromatic drive at $\omega_\text{ex}$ the steady-state amplitudes have the form $\alpha_{i,j}(t_n) \approx A_{i,j}(\omega_\text{ex})\exp^{-\ii \omega_\text{ex} t_n}$, where $A_{i,j}$ are the oscillation amplitudes. The total intensity is obtained as the sum of the oscillation amplitudes $|A_{i,j}|^2$ of all the pendula $I(\omega_\text{ex})=\sum_{i,j} |A_{i,j}(\omega_\text{ex})|^2$. The response spectra describe the total intensity as a function of the detuning $\Delta \omega=\omega_\text{ex}-\bar{\omega}$. 

Depending on the value of $\Delta\omega$ and the position $(i_p,j_p)$ of the driven site, different behaviours are expected. For $\Delta\omega$ within a band of the bulk HH model and $(i_p,j_p)$ located far from the edges, the bulk of the system is excited and so the response is dominated by delocalized band states. When $\Delta\omega$ belongs to a gap between two energy bands, the total oscillation is suppressed unless $(i_p,j_p)$ is located close to an edge of the system, so that edge states can be excited. A detailed study of these regimes will be the subject of the next Subsection.

To facilitate the reader in following the discussion, it is useful to first review some fundamental facts about the HH model. In the energy-flux plane, the states organize themselves in a recursive and self-similar structure, known as the Hofstadter butterfly. The structure of the bands and the gaps is the simplest when the magnetic flux is a rational number $\theta=p/q$ with $p,q$ being co-prime positive integers~\cite{Hofstadter}. In this special case, one has exactly $q$ sub-bands separated by $q-1$ energy gaps. When $q$ is even, the two middle bands touch at Dirac points.

The eigenstates making up the Hofstadter bands are associated with a non-trivial topological invariant, called the Chern number $C$. This invariant is related to the transverse conductivity of the integer quantum Hall effect~\cite{Thouless}. In a finite system with non-zero $C$, unidirectionally propagating states exist at the edges of the system at energies located in the gaps between bands. Since there are no states available for backscattering, these states are topologically protected from disorder~\cite{Hasan}. In most optical \cite{Hafezi1,Hafezi2,Rechtsman} and mechanical \cite{Huber,Vitelligyro} experiments so far, the presence of such chiral edge states represents the smoking gun of the non-trivial topology. We now show that such edge states can be identified also in our scheme.

\subsection{Spectra and topological edge states}
\begin{figure}[t]
\includegraphics[width=0.48\textwidth]{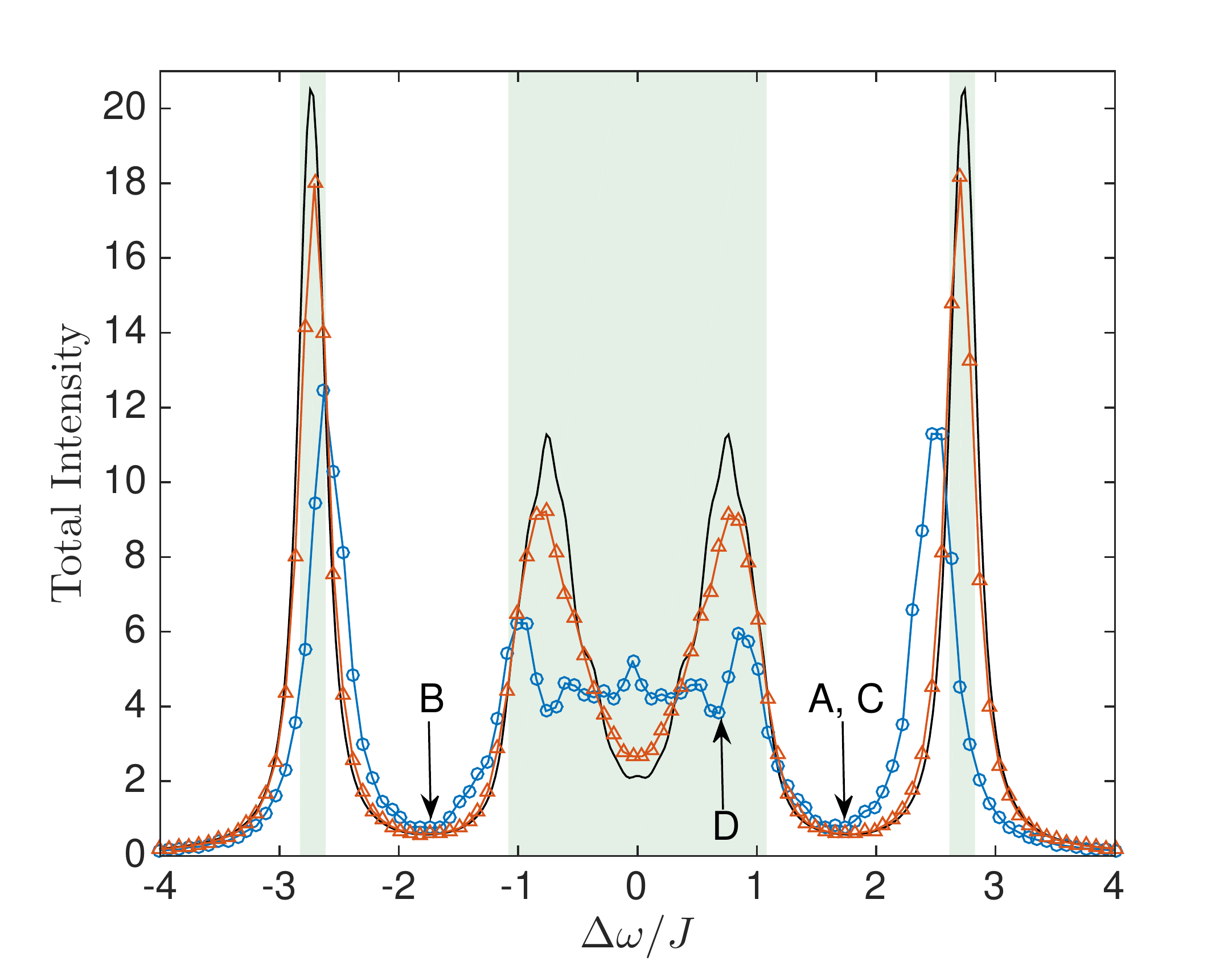}
\caption{Response spectra of a $25\times 25$ lattice with a flux per plaquette $\theta=1/4$ and losses $\gamma/J=0.1$ where the central lattice site is pumped. The blue circles and orange triangles correspond to the spectra calculated from the steady state of Eq.~\eqref{alphadrivendot} as a function of the detuning from the external driving frequency $\Delta\omega= \omega_\text{ex}-\bar{\omega}$, including the shift in Eq.~\eqref{shift}. Parameters are: $\omega_0/J=200$, $w/J=20$, $I_0=0.5$ for the blue circles, and $\omega_0/J=2000$, $w/J=50$, $I_0=0.5$ for the orange triangles. Black curve shows the spectra as obtained from the driven-dissipative HH model~\cite{Onur,Ozawa}. Green areas show the position of the HH bands. Frequencies indicated by arrows are used in Fig.~\ref{fig:topologicalstate}.}
\label{fig:spectra}
\end{figure}

The response spectra presented in Fig.~\ref{fig:spectra} are calculated by driving the central site of a lattice of $25\times 25$ pendula with a flux per plaquette of $\theta=1/4$ in units of the flux quantum. Losses are chosen to be $\gamma/J=0.1$, which is large enough to ensure that the excitation does not reach the edges of the system and that the response is only determined by the bulk properties. The position of the HH bands is highlighted by the green shaded area. The accuracy of the Floquet effective dynamics is assessed by comparing the numerical result with the prediction of the driven-dissipative HH model studied in~\cite{Onur,Ozawa}.

The blue circles show the response spectrum of a system of bare frequency $\omega_0/J=200$, temporal modulation frequency $w/J=20$ and amplitude such that $I_0=0.5$ from Eq.~\eqref{argument}. This corresponds to $V\approx 7.1$, $\Omega_x\approx 4.4$ and $\Omega_y\approx 1.1$.
The orange triangles are instead calculated for $\omega_0/J=2000$, $w/J=50$, and again $I_0=0.5$, which corresponds to $V\approx 17.7$, and the same $\Omega_x$ and $\Omega_y$ as before.
For the latter set of parameters, the inequality in Eq.~\eqref{inequality} is better satisfied and indeed the peak amplitudes of the pendula-spectra agree very well with the driven-dissipative HH model depicted with a solid black line.
With the former set of parameters, the two central peaks are difficult to resolve but the external regions of high intensity still coincide well with the position of the HH bands.
For even smaller values of $\omega_0$ the deviation from the HH model would be much more dramatic, as the RWA is no longer valid and the contribution of the counter-rotating-wave terms is not negligible any more.

\begin{figure}[t]
\hspace*{-1em}
\includegraphics[width=0.24\textwidth]{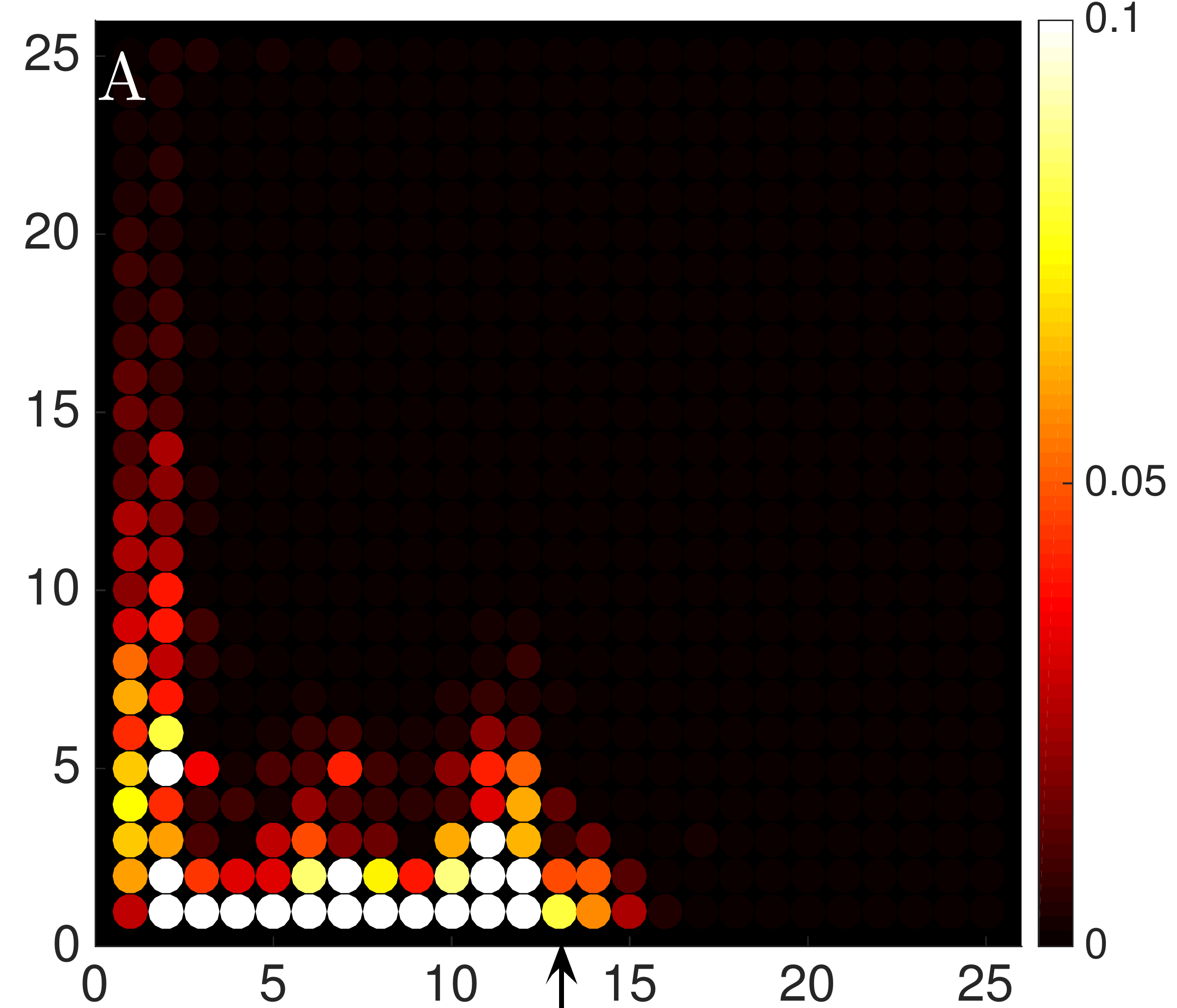}
\includegraphics[width=0.24\textwidth]{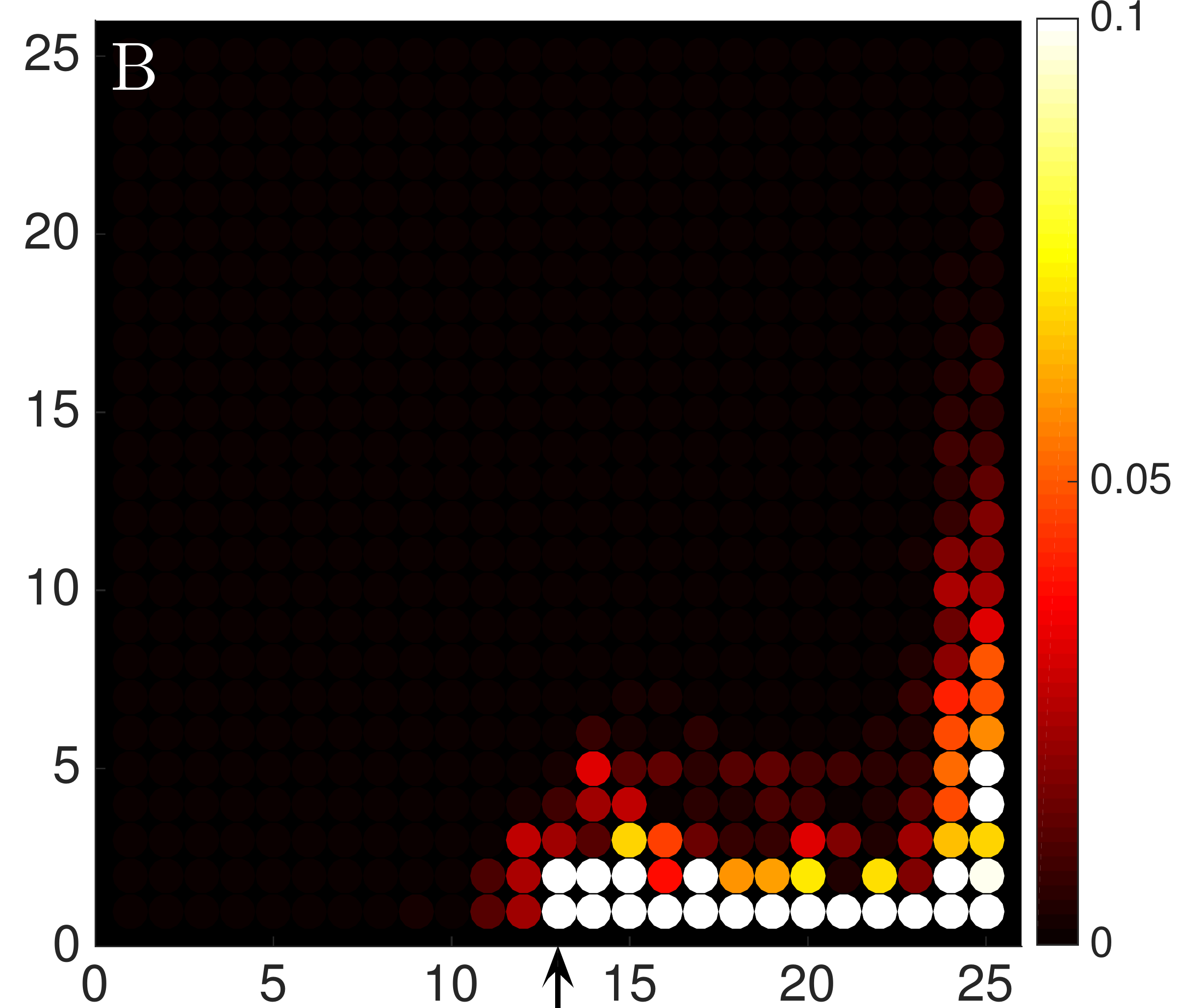}
\hspace*{-1em}
\includegraphics[width=0.24\textwidth]{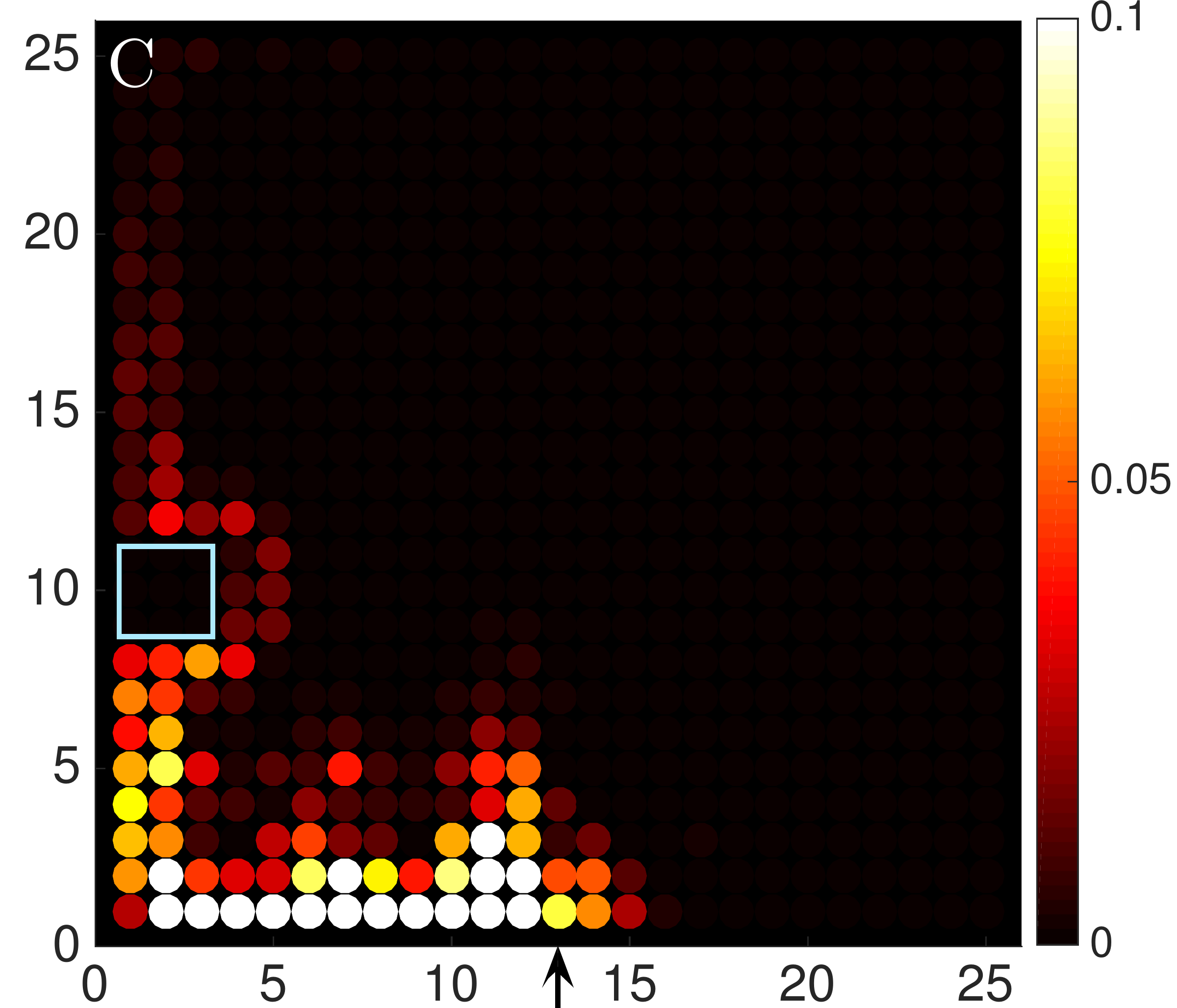}
\includegraphics[width=0.24\textwidth]{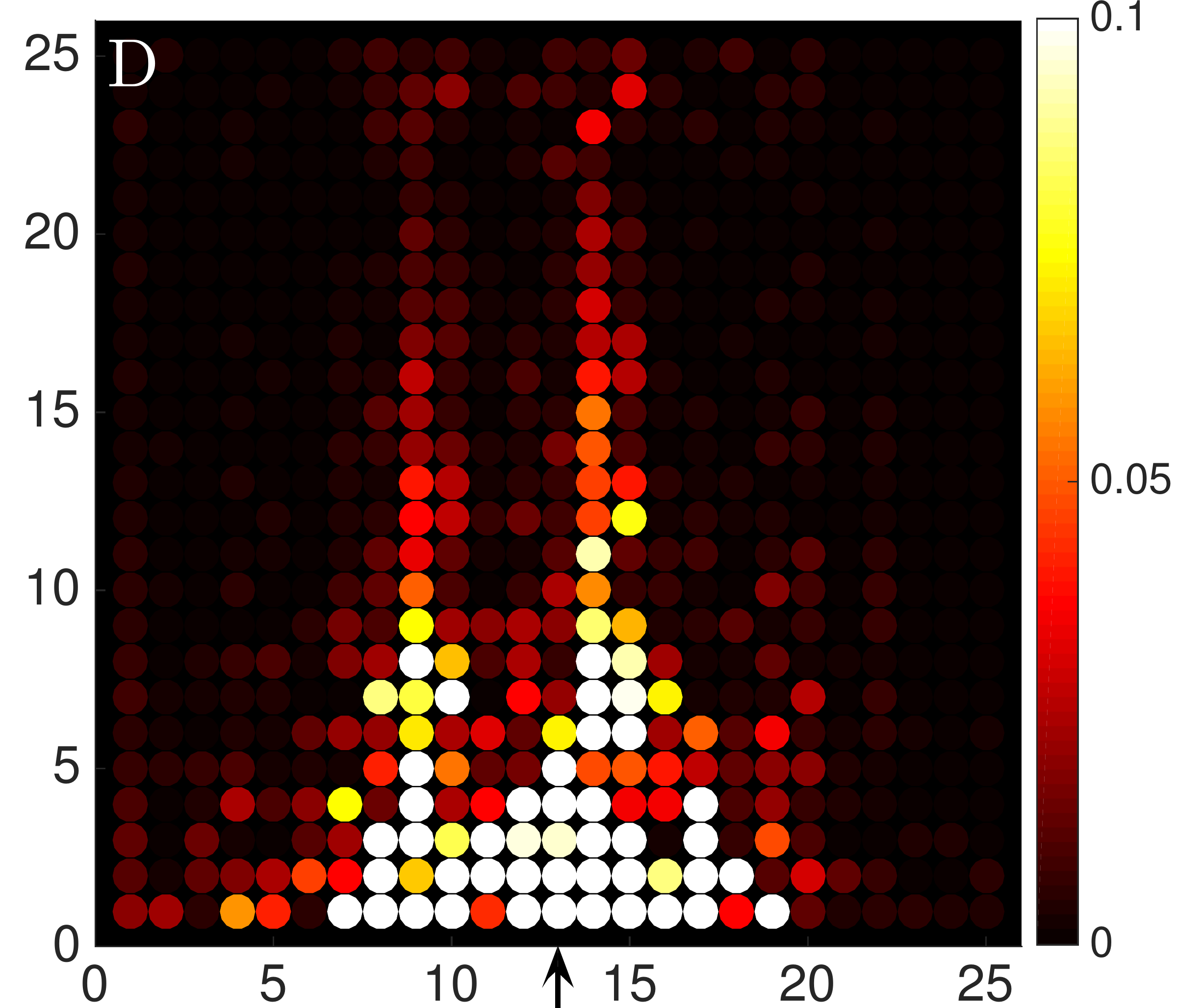}
\caption{Spatial steady-state oscillation intensity distribution for a lattice of $25\times 25$ pendula with $\theta=1/4$, $\gamma/J=0.05$, $I_0=0.5$, $w/J=20$ and $\omega_0/J=200$, obtained by pumping a single site on the lower edge for different values of the detuning of the pumping frequency $\Delta\omega/J$. Panel \textbf{A} and \textbf{C} are for $\Delta\omega/J=1.7$ located in a gap and show the excitation of an edge state with clockwise chirality. In panel \textbf{C}, the blue square indicates the position of a defect composed of $3 \times 3$ missing pendula on the left edge. Panel \textbf{B} is for $\Delta\omega/J=-1.7$ located again in a gap and shows the excitation of an edge state with counter-clockwise chirality. Panel \textbf{D} is for $\Delta\omega/J=0.7$ located in an energy band and shows the excitation of bulk states. The arrow indicates the position of the pumped site on the lower edge.}
\label{fig:topologicalstate}
\end{figure}

In Fig.~\ref{fig:topologicalstate} we show the steady state intensity distribution obtained by pumping one site in the middle of the bottom edge with different values of the detuning $\Delta\omega/J$ corresponding to the arrows in Fig.~\ref{fig:spectra}. System parameters are $\omega_0/J=200$, $w/J=20$ and $I_0=0.5$ as for the blue spectra of Fig.~\ref{fig:spectra}, while losses $\gamma/J=0.05$ are smaller in order to allow the steady state amplitude oscillations to propagate over an adequate distance. 

The intensity distributions shown in panels \textbf{A}, \textbf{B} and \textbf{C} are obtained by driving the system in the band gap where the chiral edge-states are expected. The pendula with the biggest oscillation amplitudes are indeed localized on the edge of the system. As one can see by comparing panels \textbf{A} and \textbf{B}, the direction of unidirectional propagation changes when the sign of $\Delta\omega$ is changed, as expected from the HH model. The distance covered along the edge before decaying is set by the ratio of the group velocity of the edge state over the loss rate, as usual in such driven-dissipative systems~\cite{Carusotto}. 
The topological robustness of such edge states is highlighted in panel \textbf{C}, where a defect modelled as $3\times 3$ missing pendula was introduced on the left edge of the lattice. The edge state is found to propagate around the defect, without being scattered into the bulk.
Finally, panel \textbf{D} shows the steady-state intensity distribution when the detuning corresponds to a HH band: in this case, the excitation penetrates into the bulk and shows no preferred chirality.

\subsection{Bulk wave functions}
\begin{figure}[t]
\hspace*{-1em}
\includegraphics[width=0.24\textwidth]{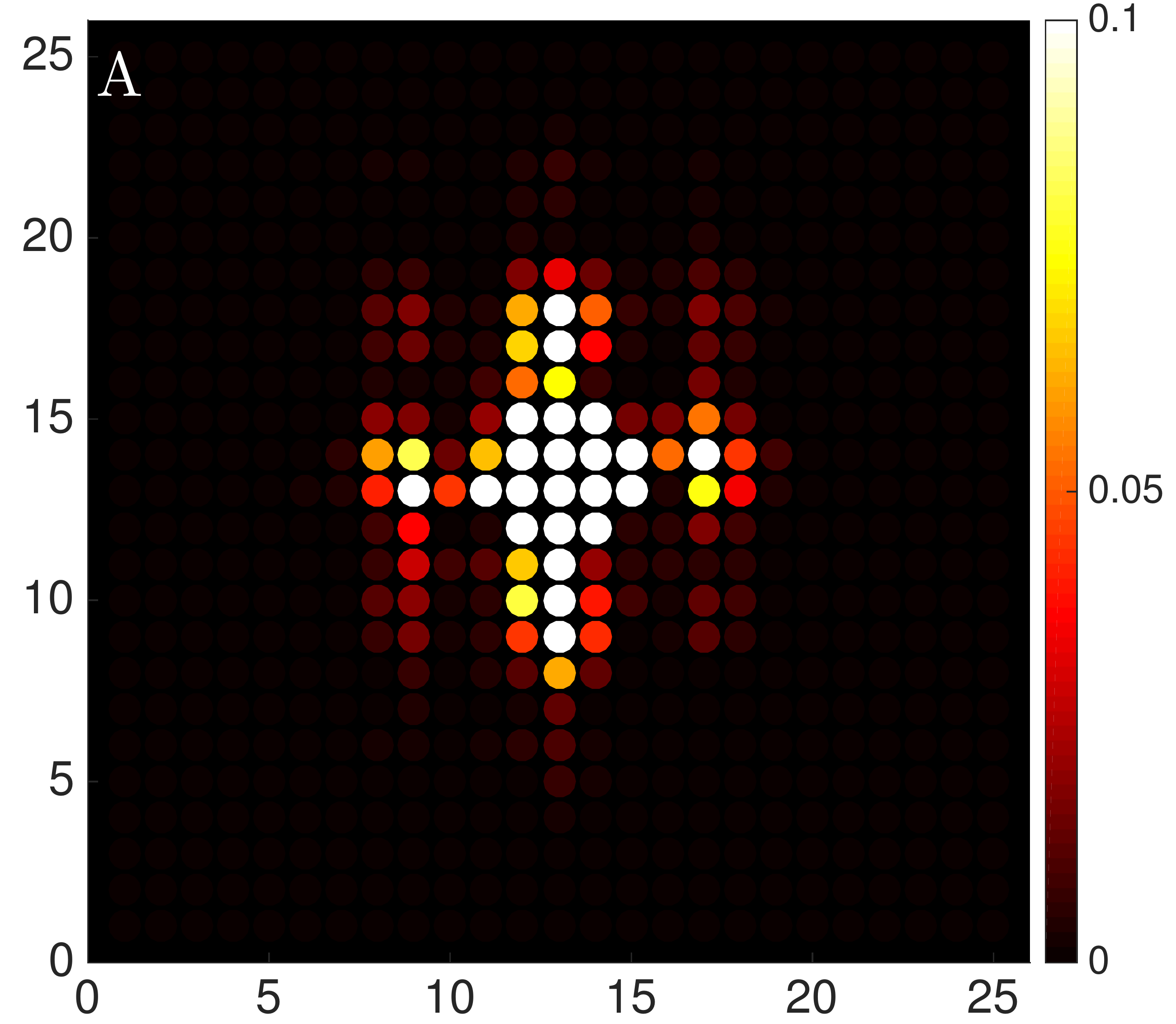}
\includegraphics[width=0.24\textwidth]{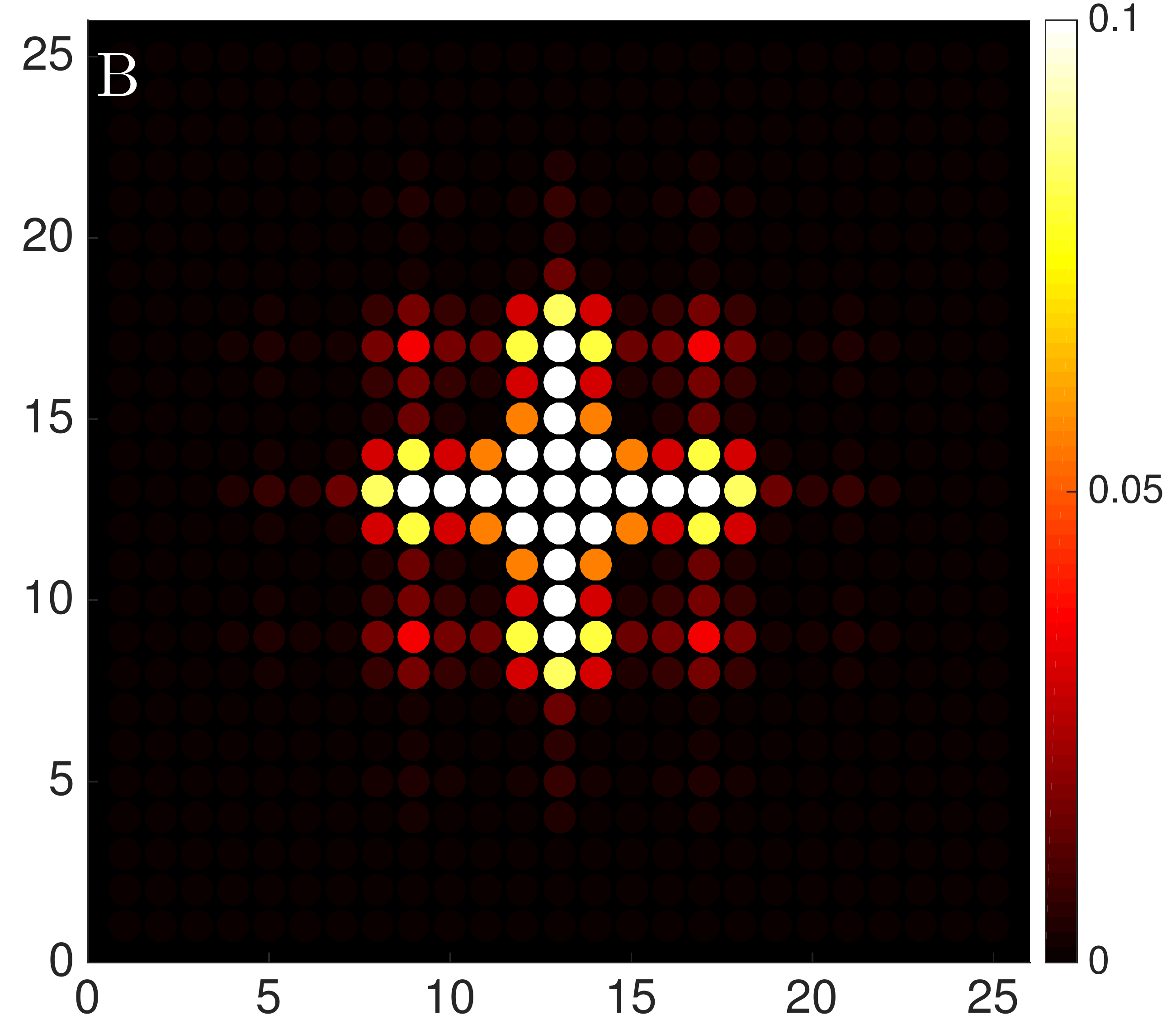}\\
\hspace*{-1em}
\includegraphics[width=0.24\textwidth]{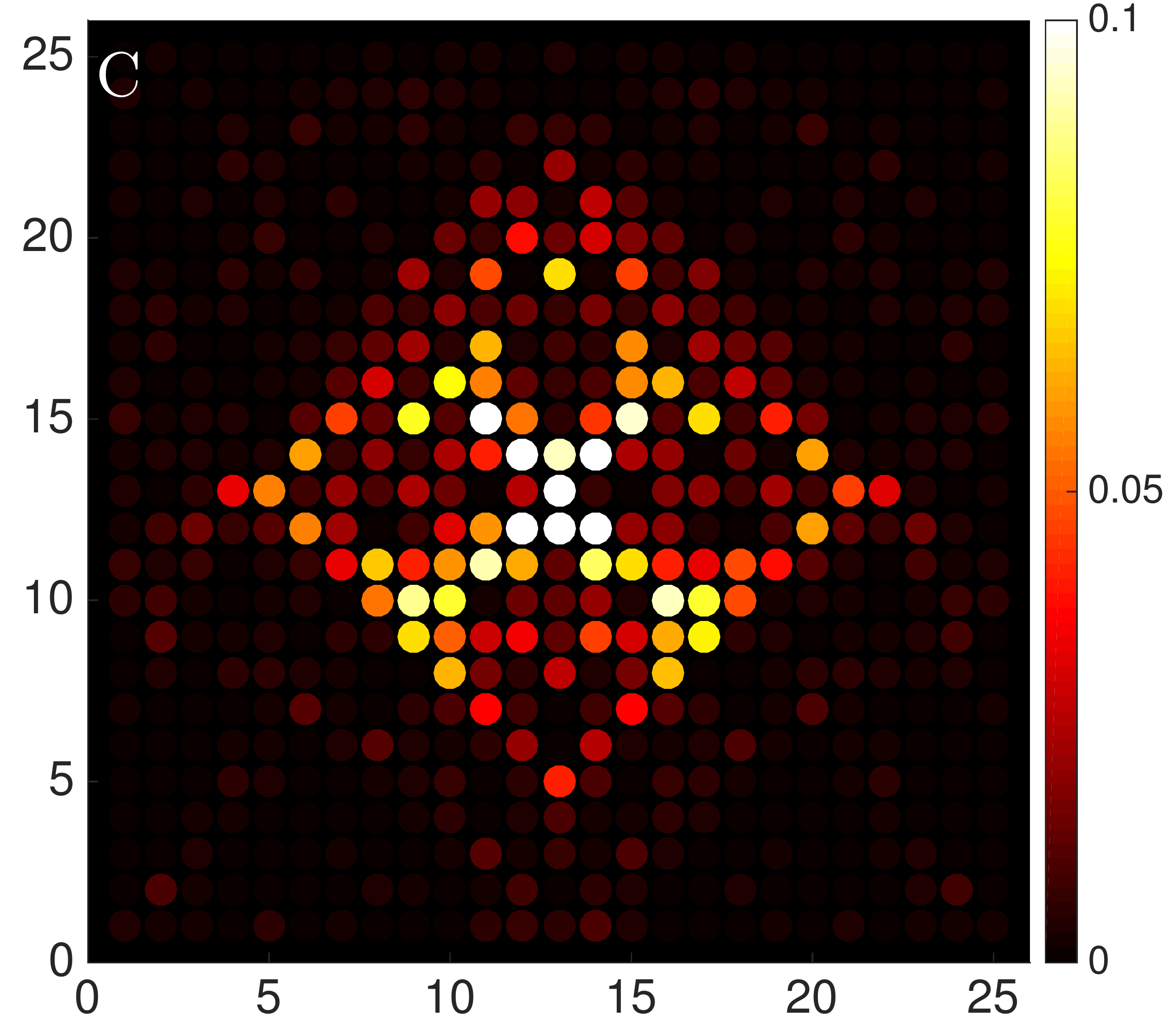}
\includegraphics[width=0.24\textwidth]{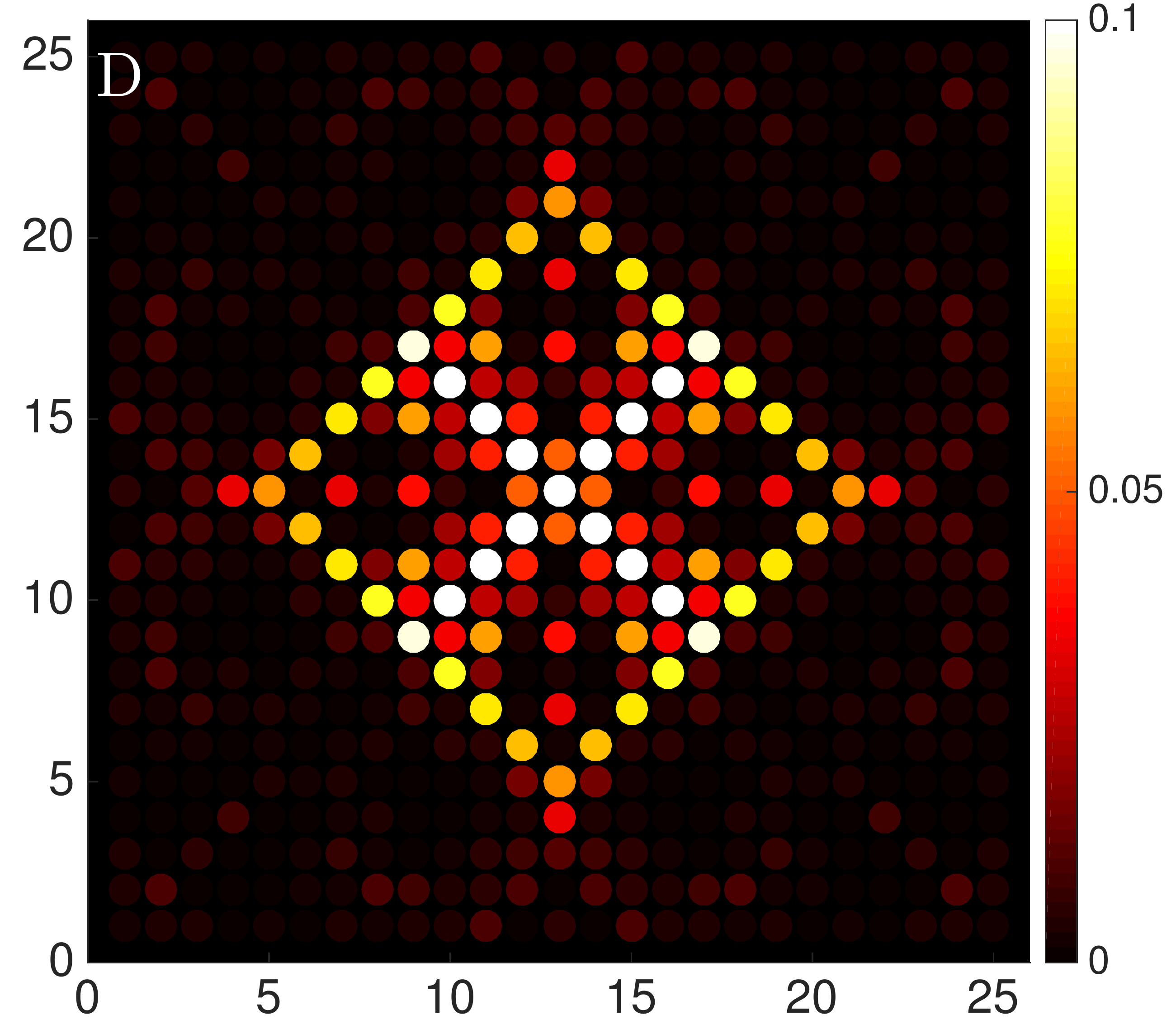}
\caption{Spatial steady-state oscillation intensity pattern for a lattice of $25\times 25$ pendula with $\gamma/J=0.1$, $I_0=0.5$, $w/J=50$ and $\omega_0/J=2000$, obtained by pumping the central site.  Panels \textbf{A} and \textbf{C} are obtained from full numerical integration of Newton's equation of motion, while panels \textbf{B} and \textbf{D} are obtained from a driven-dissipative HH model as in \cite{Onur,Ozawa}, and are shown for comparison. The first row is for $\theta=1/4$ and $\Delta\omega/J= 2.7$ corresponding to the center of the highest band. The second row is for $\theta=1/9$ and $\Delta\omega/J=0$ corresponding to the middle of the central band.}
\label{fig:steadystate}
\end{figure}

While the existence of the chiral edge states is one hallmark of a quantum Hall system, we now demonstrate that characteristic features of the driven-dissipative HH model can also be observed in the response of the bulk of the system. In Fig.~\ref{fig:steadystate}, we present the intensity distribution of the system when the central site of the lattice is driven with a detuning frequency resonant with an energy band. For the panels in the upper row, the magnetic flux is $\theta=1/4$, while $\theta=1/9$ for the panels in the lower row; losses are $\gamma/J=0.1$ in all panels. Panels \textbf{A} and \textbf{C} show the distributions of the full numerical integration with $\omega_0/J=2000$, $w/J=50$, $I_0=0.5$ and can be compared with panels \textbf{B} and \textbf{D} that are obtained from the driven-dissipative HH model as in \cite{Onur,Ozawa}. We notice that the two pattern distributions are in excellent qualitative agreement, in particular where the periodicity of the pattern is concerned. As expected, when the natural frequency $\omega_0$ is reduced away from the RWA limit, the spatially symmetric structure of the oscillation amplitude pattern is broken and the agreement with the driven-dissipative HH model is less good (not shown). 

\subsection{The Hofstadter butterfly}
\begin{figure}[t]
\includegraphics[width=0.48\textwidth]{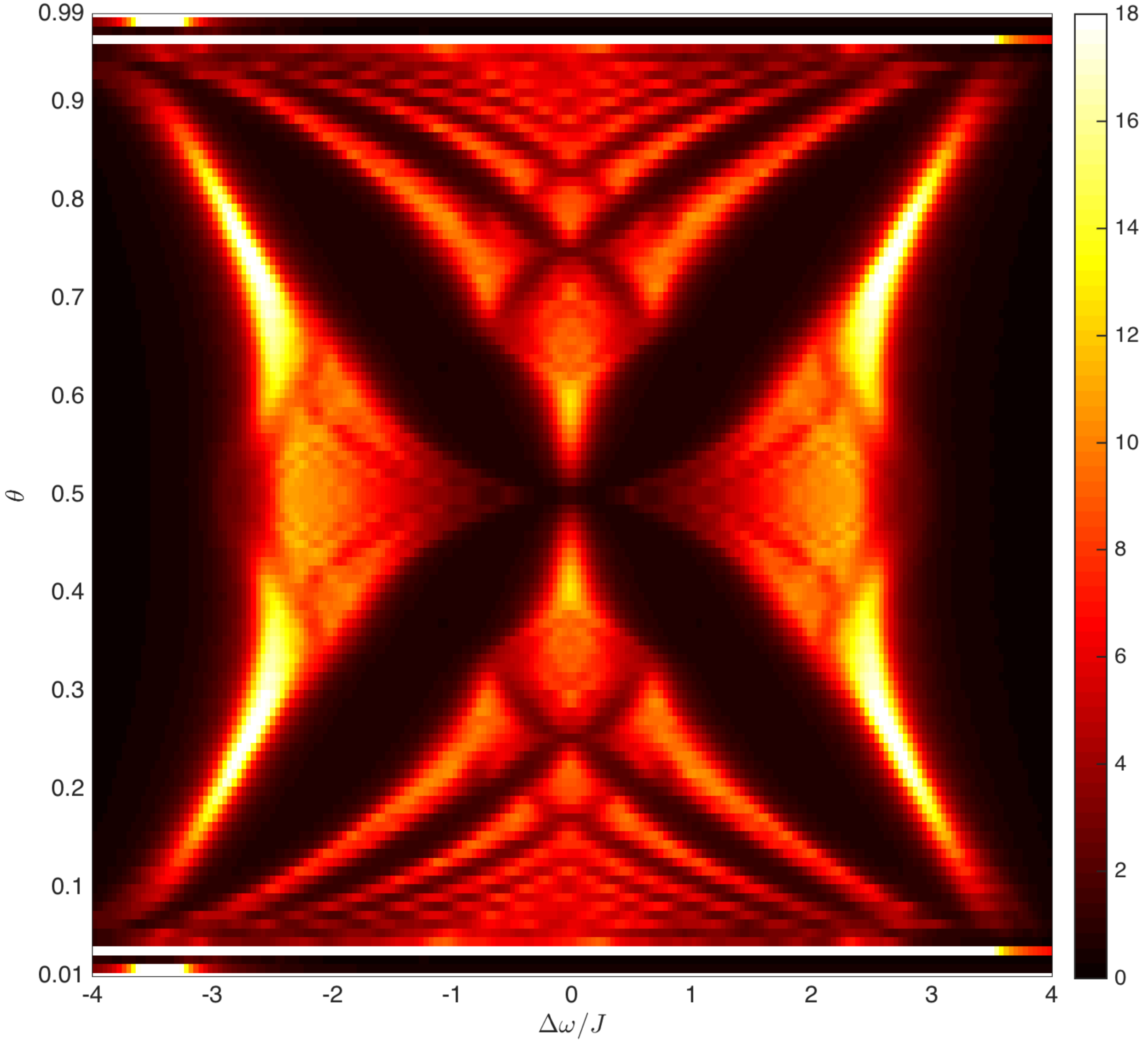}
\caption{Total intensity response spectra as a function of flux per plaquette $\theta$ and frequency detuning $\Delta \omega/J$. The total intensity of the oscillation is depicted with the color scale, where lower intensities are in darker colors. Black regions indicating negligible intensities correspond to the energy gaps between energy bands. The peculiar structure of the Hofstadter butterfly pattern is clearly visible. System parameters are: $\omega_0/J=2000$, $w/J=50$, $\gamma/J=0.1$, $I_0=0.5$, for a lattice of $25\times 25$ sites where only the central one is pumped.}
\label{fig:butterfly}
\end{figure}

The scheme that we have proposed in the previous Sections allows us to change the magnetic flux $\theta$ per plaquette by simply adjusting the phase of the modulation of each pendula in Eq.~\eqref{timemodulation}. 
In this way, our scheme is sufficiently flexible to explore a wide portion of the flux-frequency plane. Collecting a large number of numerically calculated spectra of the total intensity for different values of $\theta$ in a single color-plot, we obtain the pattern shown in Fig.~\ref{fig:butterfly}. Although the presence of dissipation does not allow us to resolve details of the fractal structure that are smaller than the loss rate, the emerging picture shows a close resemblance to the Hofstadter butterfly~\cite{Hofstadter}. Such a result would constitute the first experimental visualization of the Hofstadter butterfly in classical physics.

In order to speed up the numerical calculations needed to plot Fig.~\ref{fig:butterfly}, the equations of motion in Eq.~\eqref{alphadrivendot} have been solved by expanding the solution in the Fourier basis:
\begin{equation}
\alpha_{i,j}(t)=\sum_{m=-M}^M \sum_{n = \pm 1} \alpha^{(m,n)}_{i,j} \e^{\ii m w t}\e^{\ii n \omega_\text{ex}t} 
\end{equation}
and then truncating the series to a value $M$ that is sufficiently large to ensure the convergence of the solution. Solving the linear system for the Fourier components coupled by the time-dependent modulation is equivalent to, but much faster than, solving the full differential equations until reaching the steady state. More details on this decomposition method are found in Appendix~\ref{app:Fourier}.

It is interesting to note that the agreement with the well-known Hofstadter butterfly gets worse in the top and bottom regions of the spectra and eventually breaks down. Those areas correspond to very low values of the magnetic flux $\theta$ for which the number of bands is large and the smoothing has most dramatic consequences. 
To understand the break-down from the butterfly structure, we checked the solutions with the full numerical Runge-Kutta integration method and we observed that in these regions the system is dynamically unstable. This means that a stationary steady state can not exist and therefore the Hofstadter butterfly pattern completely breaks down.
These deviations stem from the strength of the counter-rotating-wave-terms as controlled by the amplitude of the temporal modulation $V$. In our simulation, we take $I_0 = 0.5$ to be fixed which, from Eq.~\eqref{argument}, results in an increase of $V$ as the flux is decreased.
Larger $V$ implies a larger contribution from counter-rotating-wave terms, that results in the emergence of a parametric instability~\cite{Arnold}. 
We note that the spectra obtained from the Fourier decomposition method appear to be still stable despite the exponential growth of the Runge-Kutta solution because such a method begins from the ansatz of a steady-state solution and so cannot predict the instability. More insights into the instability will be given later in the paper.

\subsection{Integer quantum Hall effect and Chern number}
In the previous Sections, we have demonstrated that the system of frequency-modulated coupled harmonic oscillators can be a classical simulator of the HH model. As a final point, we now show that an analogue of the integer quantum Hall effect (IQHE) can be observed even in a classical mechanical context. A main feature of the IQHE is that a current appears in the direction orthogonal to the applied electric field where the associated conductivity is an integer multiple of the conductance quantum $e^2/h$. This integer is related to the Chern number of the populated bands of the HH model. For the bosonic model considered here, these bands can be populated by a suitable excitation scheme. Our proposal is inspired by a recent work on photonic systems~\cite{Ozawa}.

\begin{figure}[t]
\hspace*{-1em}
\includegraphics[width=0.24\textwidth]{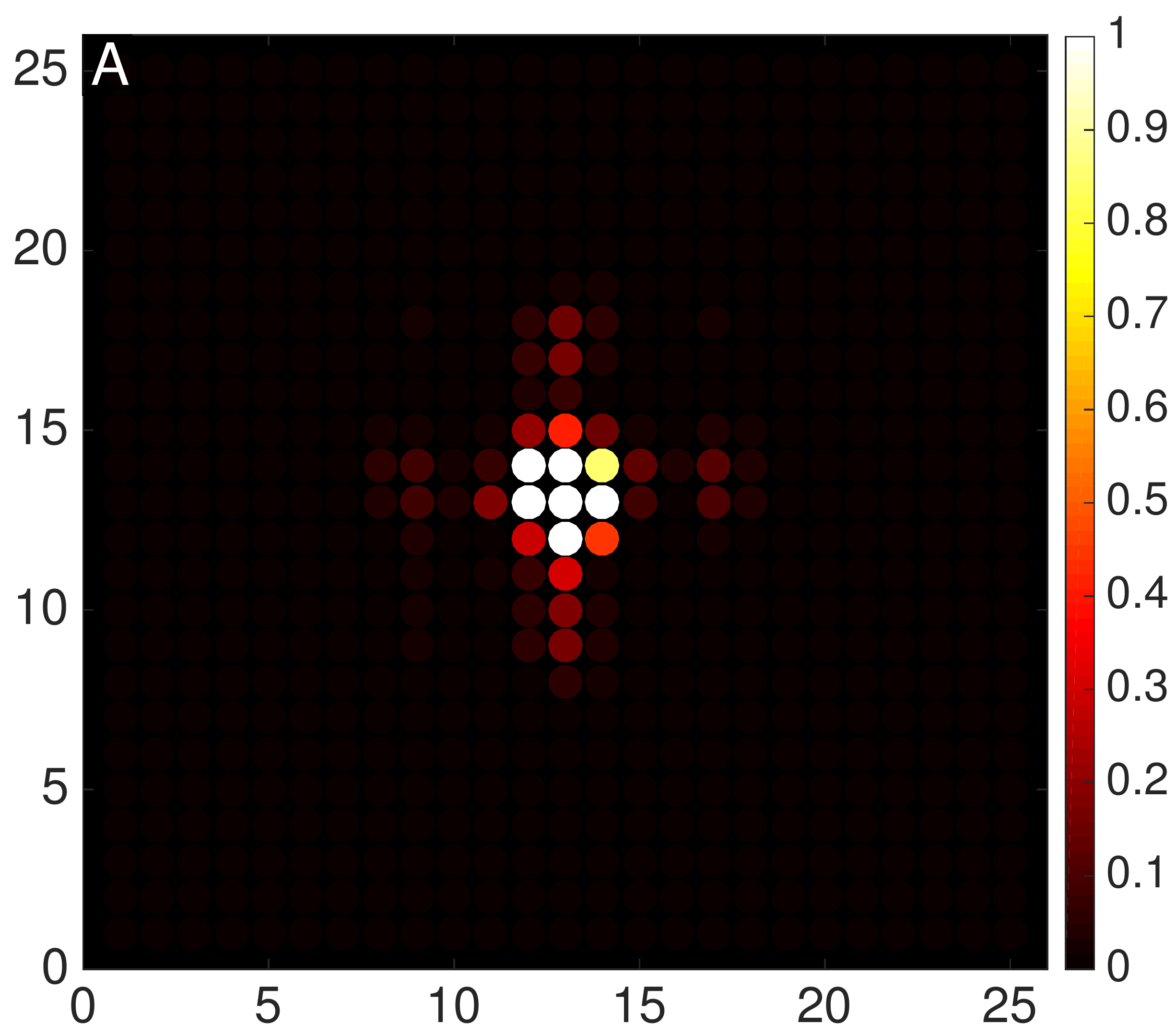}
\includegraphics[width=0.24\textwidth]{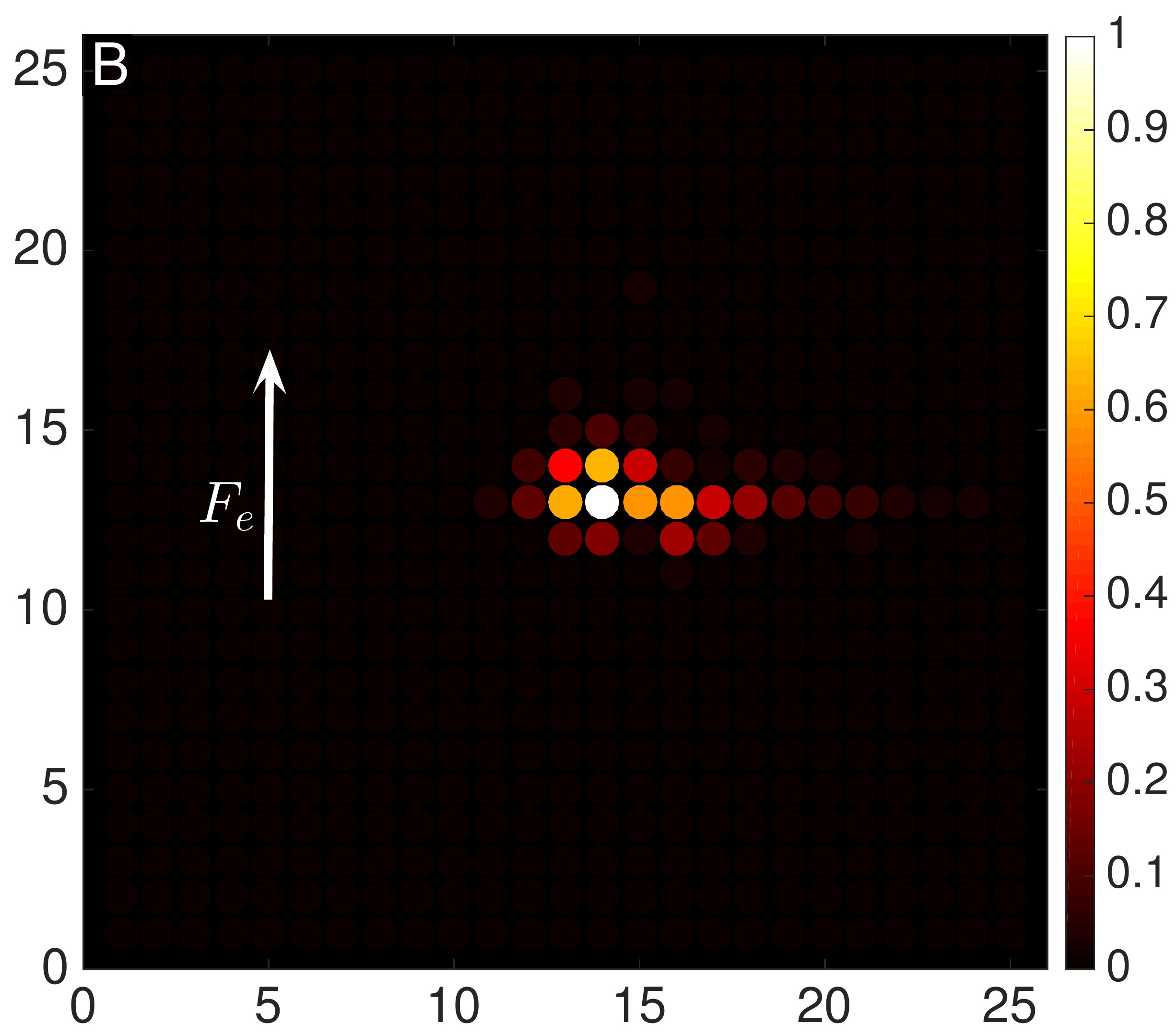}
\caption{Spatial steady-state oscillation intensity distribution for a lattice of $25\times 25$ pendula with $\theta=1/4$, $\gamma/J=0.1$, $I_0=0.5$, $w/J=50$ and $\omega_0/J=2000$, obtained by pumping the central site with a detuning corresponding to the middle of the lowest band $\Delta\omega/J=-2.7$. For panel \textbf{A} there is no external force $F_e/J=0$, and the centre of the intensity distribution indeed remains at the centre of the system. For panel \textbf{B}, the force $F_e/J=0.5$ is directed along the $y$-direction as indicated by the arrow and induces a sizeable rightward shift of the centre of the distribution towards the positive $x$ direction as expected from the IQHE.}
\label{fig:halleffect}
\end{figure}

In order to introduce an analogue of the electric field in our model, we need to generate a linear potential acting on the excitation field $\alpha_{i,j}$. Applying such a potential along the positive $y$-direction would correspond to a term of the form $-\ii F_e j \alpha_{i,j}$ in Eq.~\eqref{alphadrivendot}. Equivalently, and perhaps more straightforwardly, the analogue electric field could be introduced through an additional linear gradient of the bare oscillation frequency along the $y$-direction in Eq.~\eqref{staticmod}, with $\tilde{\omega}_{i,j} \rightarrow \tilde{\omega}_{i,j} + F_e j$. 

In Fig.~\ref{fig:halleffect}-\textbf{A} we show the intensity distribution of a system of $25\times25$ pendula with $\theta=1/4$ and $\gamma/J=0.1$ in the absence of the external force and for a detuning $\Delta\omega/J=-2.7$ such that only the lowest band is excited with a resulting population that is approximately uniform. As discussed in~\cite{Ozawa}, these conditions can be fulfilled provided the energy width of the band $\Delta_\text{width}$ is much smaller than the band gap $\Delta_\text{gap}$ and the loss rate falls in between these two energy scales. The intensity pattern is symmetric both in $x$ and in $y$ and is centred around the middle point. 
When a non-zero force $F_e/J=0.5$ along the vertical direction is applied as in panel~\textbf{B}, the oscillation pattern clearly shifts to the right in the direction orthogonal to the applied force. This is a manifestation of the Lorentz force and a clear signature of a Hall-like effect.

In order to assess the quantized nature of this Hall effect, we can quantify the lateral shift as $\langle x \rangle \equiv \sum_{i,j} j |A_{i,j}|^2/I$ and compare it to the prediction for the integer quantum Hall effect in driven-dissipative systems~\cite{Ozawa}, 
\begin{equation}
\langle x\rangle= -F_e \left( \frac{q C}{2 \pi \gamma}+ \eta \right)
\label{hallshift}
 \end{equation}
where $q$ is the denominator of the rational flux $\theta=p/q$ and $C$ is the Chern number of the excited band. The real number $\eta$ quantifies the (spurious) contribution of the bandwidth, that is responsible for a non-uniform population of the band, as well of the other neighbouring bands. The quantized nature of the IQHE is most apparent in the small $\gamma$ limit in the coefficient of the leading term.

\begin{table}[t]
\centering
  \begin{tabular}{| m{4em}  m{4em} | m{4em} | m{4em} | m{4em} |}
    \hline
 Band  & & $1^\textrm{st}$ & $2^\textrm{nd}$ & $3^\textrm{rd}$  \\
    \hline
    \hline	
  {\multirow{2}{*}{$\theta=\frac{1}{3}$} } & \multicolumn{1}{ |c| }{$C_e$} & $-0.803$ &  & $-0.873$\\
    \cline{2-5}  \multicolumn{1}{ |c }{} & \multicolumn{1}{ |c| }{$C$} & $-1$ & $+2$ & $-1$\\
      \hline\hline
       {\multirow{2}{*}{$\theta=\frac{1}{4}$} } & \multicolumn{1}{ |c| }{$C_e$} & $-0.972$ &  & \\
    \cline{2-5}  \multicolumn{1}{ |c }{} & \multicolumn{1}{ |c| }{$C$} & $-1$ & $+1$ & $+1$\\
      \hline\hline
        {\multirow{2}{*}{$\theta=\frac{1}{5}$} } & \multicolumn{1}{ |c| }{$C_e$} & $-0.878$ &  & \\
    \cline{2-5}  \multicolumn{1}{ |c }{} & \multicolumn{1}{ |c| }{$C$} & $-1$ & $-1$ & $+4$\\
      \hline\hline
        {\multirow{2}{*}{$\theta=\frac{2}{5}$} } & \multicolumn{1}{ |c| }{$C_e$} &  &  & $+1.725$\\
    \cline{2-5}  \multicolumn{1}{ |c }{} & \multicolumn{1}{ |c| }{$C$} & $+2$ & $-3$ & $+2$\\
      \hline
  \end{tabular}
  \caption{Estimated value $C_e$ of the Chern numbers as calculated from Eq.~\eqref{hallshift} for a system of $25\times 25$ lattice sites with the parameters given in the main text. These are compared to the exact values $C$, for different bands and several values of the rational flux $\theta$.}
  \label{table:Chern}
\end{table}

In Table~\ref{table:Chern} we summarize the estimated Chern numbers $C_e$ of different bands for four different $\theta$ as calculated from Eq.~\eqref{hallshift} for a system of $25\times 25$ lattice sites, $\omega_0/J=2000$, $w/J=50$, $F_e/J=0.07$ and $\Delta_\text{width}<\gamma<\Delta_\text{gap}$. The constant $\eta$, that does not depend on the losses $\gamma$, is eliminated by calculating the linear coefficient in the shift given in Eq.~\eqref{hallshift} for two values of $\gamma \in [\Delta_\text{width},\Delta_\text{gap}]$. 
Choosing any value in this frequency range only affects the calculation of the Chern number to the least-significant digit.
When the method is not applicable because $\Delta_\text{width} \gtrsim \Delta_\text{gap}$, the corresponding Chern number of these bands was not calculated. In all other cases, the agreement with the Chern number $C$ of the HH model is good and confirms the expected integer quantum Hall effect in a classical mechanics context.

\section{Parametric instability}
\label{sec:instability}

In this Section we discuss in more detail the instability observed in the small flux region of the Hofstadter butterfly of Fig.\ref{fig:butterfly} and the corrections to the HH model due to the counter-rotating-wave terms in the equations of motion Eq.~\eqref{alphadrivendot}. As the amplitude of these terms is proportional to $V/\omega_0$, their importance for a fixed $I_0$ increases as the magnetic flux $\theta$ is lowered according to Eq.~\eqref{argument}, explaining why the instability was only observed close to the upper and lower edges of the butterfly pattern.

In order to more easily access this physics, we have introduced a cleaner strategy to reinforce the counter-rotating-wave terms by decreasing the bare frequency of the pendula $\omega_0$ while keeping other parameters fixed. This approach comes with the great advantage that we can compare energy spectra with the same value of the magnetic flux as the strength of the counter-rotating-wave terms increases.
In the following we allow the bare frequency to be $\omega_0 \approx s w \gg \Omega_{x,y}$, by relaxing the constraint in Eq.~\eqref{inequality}. 

Recently, there have been several works discussing the role of the counter-rotating-wave terms in bosonic Hamiltonians on topology~\cite{Murakami,Peano,Brandes,Liew}. However, exception made for a work on edge state instabilities~\cite{Barnett}, only a little attention has been so far devoted to the interplay between topology and parametric instabilities. We now investigate how counter-rotating-wave terms lead to the appearance of complex energy modes in the spectrum and to a dramatic distortion of the HH bands which is accompanied by a disappearance of topological edge states.

To this end, it is convenient to recast the equations in Eq.~\eqref{alphadrivendot}, together with the corresponding equations for $\alpha^*$, without driving and dissipation, in the matrix form:

\begin{equation}
\ii \frac{\partial}{\partial t} 
\begin{pmatrix}
\vec{\alpha}\\
\vec{\alpha}^*
\end{pmatrix}=
\mathcal{M}(t)
\begin{pmatrix}
\vec{\alpha}\\
\vec{\alpha}^*
\end{pmatrix}
\label{bogo}
\end{equation}
where $\vec{\alpha}= (\alpha_{1,1} \ldots \alpha_{N_x, N_y})^\top$ is a column vector. The eigenvalues of the $2N\times 2N$ time periodic matrix $\mathcal{M}(t)$ represent the oscillation frequencies of the system. Due to the specific form of the coupling between the variables $\vec{\alpha}$ and $\vec{\alpha}^*$, we notice that this matrix is not Hermitian with respect to the standard positive-definite inner product. Its block-structure:
\begin{equation}
\mathcal{M}(t)=
\begin{pmatrix}
\quad A(t) & \quad B(t)\\
-B(t) & -A(t)
\end{pmatrix}
\end{equation}
where $A(t)$ and $B(t)$ are $N\times N$ real symmetric matrices, periodic with the modulation period $T=2\pi/w$, guarantees that it is Hermitian with respect to the indefinite-inner product defined by the matrix:
\begin{equation}
\eta=\begin{pmatrix} 1 &0 \\ 0 & -1 \end{pmatrix},
\end{equation}
and of common use in the Bogoliubov theory of dilute Bose gases~\cite{castin, leonhardt}. The non-Hermiticity of $\mathcal{M}(t)$ with respect to the standard inner product is crucial for the appearance of the complex energy-modes and the emergence of the instability.

To treat the time-dependent problem, we can apply the Floquet theory of time-periodic systems. The evolution within one period $T$ of the modulation is then expressed as that with a time-independent stroboscopic Floquet matrix $\mathcal{M}_F$, defined as the time-ordered exponential \cite{Bukov}:
\begin{equation}
U= \e^{-\ii \mathcal{M}_F T} \equiv \mathcal{T} \e^{-\ii \int_{0}^{T} \mathcal{M}(t) \mathrm{d}t}.
\label{timeevolution}
\end{equation}

If $\mathcal{M}(t)$ were a Hermitian matrix, then $U$ would be unitary and the eigenvalues of the stroboscopic Floquet matrix $\mathcal{M}_F$ would always be real \cite{Shirley}. 
In our system, instead, the matrix $\mathcal{M}$ describes, in terms of the complex variables in Eq.~\eqref{transformation}, the one-period evolution of the system which is symplectic in the original $z_{i,j}$ and $p_{i,j}$ variables.  As a result, this matrix is not Hermitian with respect to the standard inner product, which implies that, in general, the eigenvalues of the Floquet matrix can also be complex, so the system can be dynamically unstable \cite{leonhardt, jain}.

The matrix $U$ is the \textit{mapping at a period} of the system with periodically varying parameters \cite{Arnold}, and it is the classical analogue of the stroboscopic time-evolution operator: $ U | \psi(0) \rangle = | \psi(T) \rangle$. It is known from classical mechanics textbooks that if all the eigenvalues $\lambda_n$ of the mapping at a period $U$ are distinct and lie on the unit circle in the complex plane, the system is strongly stable \cite{Krein}. 
Since the Floquet matrix is related to $U$ by Eq.~\eqref{timeevolution}, the stability of the system can be checked just by looking at the eigenvalues of $\mathcal{M}_F$. Notice that the condition on the eigenvalues of $U$ to lie on the unit circle implies that the Floquet spectra has to be real. In fact, if $\epsilon_n$ is an eigenvalue of the Floquet matrix $\mathcal{M}_F$, then it is related to the eigenvalue of $U$ as:
\begin{equation}
\lambda_n= \e^{-\ii  \epsilon_n T}
\end{equation}
from Eq.~\eqref{timeevolution}. It is straightforward to check that if $\epsilon_n \in \mathbb{R}$, then $|\lambda_n|=1$ is on the unit circle. 

In analogy with the above-mentioned Bogoliubov theory~\cite{castin}, it directly follows from the specific form of the matrix $\mathcal{M}(t)$ that if $( \vec{u}_n, \vec{v}_n)^\top$ is an eigenvector of $\mathcal{M}_F$ with eigenvalue $\epsilon_n$, then $(\vec{v}_n^*, \vec{u}_n^*)^\top$ is also an eigenvector but with opposite eigenvalue $-\epsilon_n^*$.
The $2N$ eigenvalues of $\mathcal{M}_F$ can be then distinguished according to their \textit{Krein signature}~\cite{Arnold, Krein}:
\begin{equation}
\mathcal{K}= \text{sign}(|u_n|^2-|v_n|^2)
\label{kreinsignature}
\end{equation}
that is the sign of the norm corresponding to the $\eta$ scalar product of the above-mentioned Bogoliubov theory~\cite{castin, leonhardt}. By definition, the Krein signature is either $\pm1, 0$. All stable eigenvalues of $\mathcal{M}_F$ have either positive or negative Krein signature, while all unstable eigenvalues have zero Krein signature; this also holds for the eigenvalues of $U$. 

\begin{figure}[t]
\hspace*{-0.5em}
\includegraphics[width=0.24\textwidth]{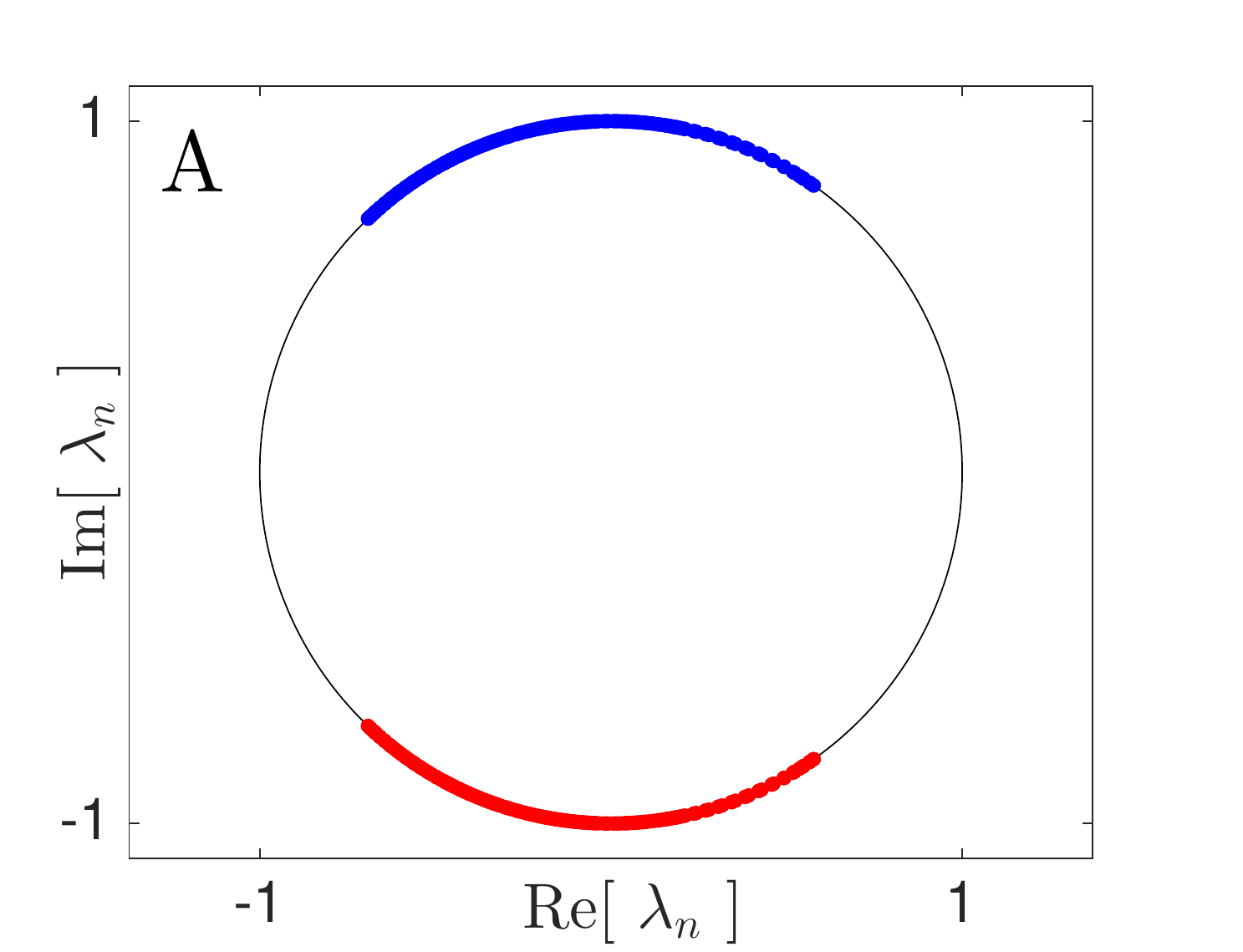}
\includegraphics[width=0.24\textwidth]{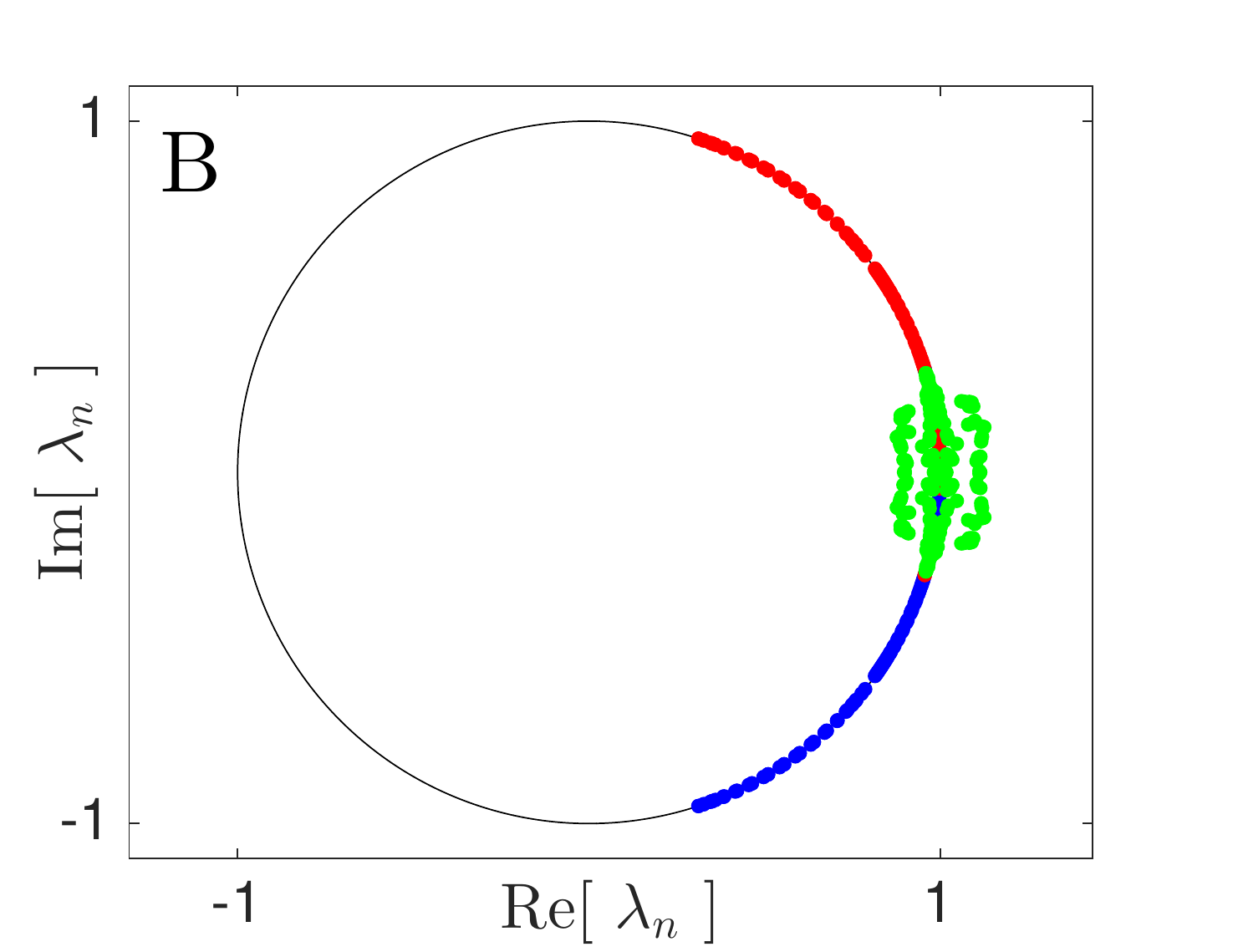}
\caption{Representation on the unit circle of the eigenvalues of the matrix $U$ in the complex plane. Red points corresponds to eigenvalues with Krein signature $\mathcal{K}>0$, points in blue are the eigenvalues with $\mathcal{K}<0$ and green points are unstable eigenvalues with $\mathcal{K}=0$. Panel \textbf{A} shows a stable configuration for  $\omega_0/w=5.4$, where all eigenvalues lies on the unit circles. Panel \textbf{B} shows the unstable configuration $\omega_0/w=5.08$, where some eigenvalues are clearly not lying on the unit circle. Parameters are: $\theta=1/5$, $s=5$, $w/J=50$.}
\label{fig:krein}
\end{figure}

In Fig.~\ref{fig:krein} we show the representation on the unit circle of the eigenvalues of the matrix $U$, coloured according to their Krein signature. Eigenvalues with positive $\mathcal{K}$ are shown in red, eigenvalues with negative $\mathcal{K}$ are displayed in blue and unstable eigenvalues with zero Krein signature are displayed in green.
As the ratio $\omega_0/w$ is changed, the eigenvalues $\lambda_n$ of $U$ rotate on the unit circle. Whenever eigenvalues with opposite Krein signature meet, they can leave the unit circle, meaning that the system becomes unstable. This instability happens every half-round-trip: since the eigenvalues $\lambda_n$ perform a full rotation on the unit circle when $\omega_0/w$ is increased by $1$, the instability occurs each time $2\omega_0/w= m$, with $m \in \mathbb{N}$.

\begin{figure}[t]
\includegraphics[width=0.5\textwidth]{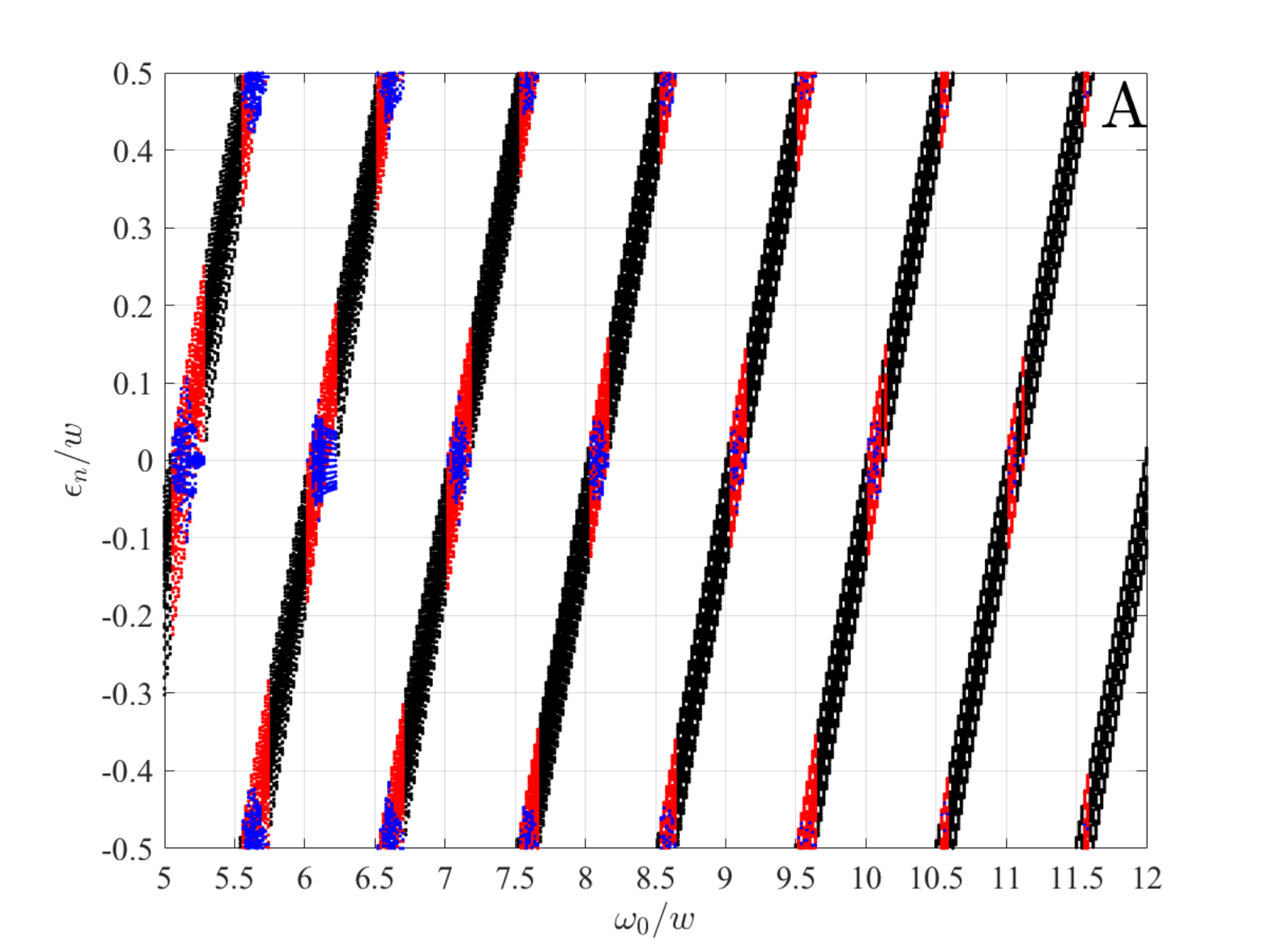}
\includegraphics[width=0.5\textwidth]{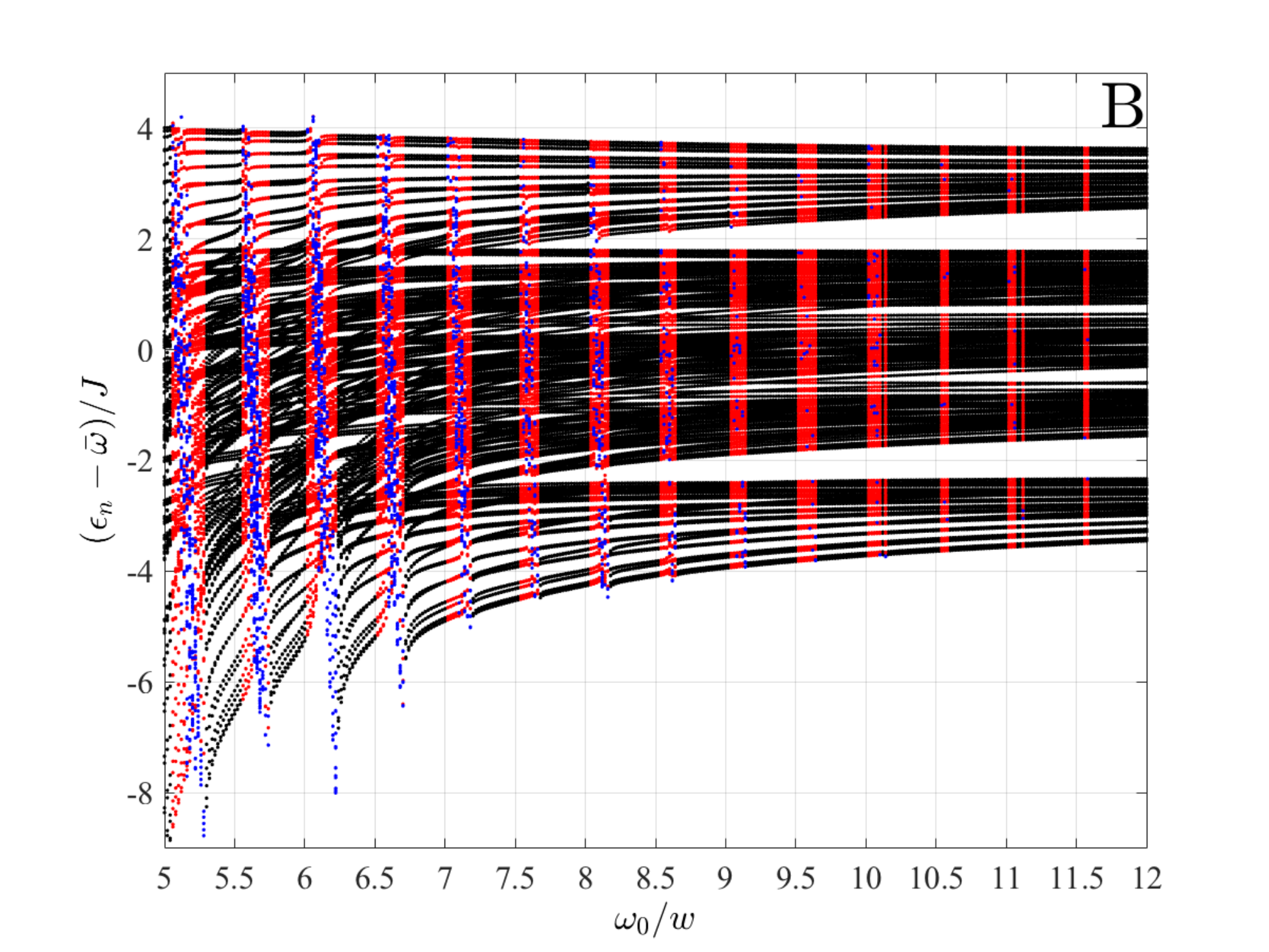}
\caption{Floquet quasi-energies $\epsilon_n$, modulo $w$, associated with either a positive Krein-signature $\mathcal{K}$ or with $\mathcal{K}=0$ and $|\lambda_n| \geq 1$. The system has a magnetic flux $\theta=1/5$, with periodic boundary conditions; driving and dissipation are not included. The energy spectra are presented as a function of $\omega_0/w$, for a fixed value of the amplitude modulation $V/w\approx 0.4$. When all the eigenvalues at a particular frequency are stable, we denote them in black, otherwise we denote them either in red or blue. Red points are purely real modes, while blue points are the unstable modes, plotted according to the real part of their complex energy. Panel \textbf{A} shows the eigenvalues  in units of $w$, while in panel \textbf{B} the same eigenvalues are shifted by $\bar{\omega}$ as defined in Eq.~\eqref{shift}. Parameters are: $\theta=1/5$, $s=5$, $w/J=50$.}
\label{fig:stability}
\end{figure}

In Fig.~\ref{fig:stability} the Floquet energy spectra $\epsilon_n$ of the system are shown as the bare frequency $\omega_0$ is changed for a fixed value of $w$. The magnetic flux is $\theta=1/5$ and periodic boundary conditions are used, so that the edge states do not appear in the spectra. With no loss of information, in the figure we only present stable eigenvalues associated with a positive Krein-signature and unstable eigenvalues with $|\lambda_n| \geq 1$.
In Panel~\textbf{A}, the eigenvalues are displayed modulo $w$, while in Panel~\textbf{B} the same eigenvalues $\epsilon_n$ are shifted by $\bar{\omega}$ as defined in Eq.~\eqref{shift}. 

When all the eigenvalues at a particular frequency are stable, we denote them in black, otherwise we denote them either in red or blue. The points in blue correspond to the modes which have become unstable, plotted here according to the real part of the energy. The points in red are eigenmodes with purely real energies. We refer to frequencies for which there are blue points as instability regions.

As anticipated above, we observe that the instability regions occur every integer and half-integer times of $\omega_0/w$. In particular, purely imaginary modes start to develop around $\epsilon_n/w =0$ and  $\epsilon_n/w =\pm 1/2$. 
As $\omega_0/w$ is increased, the size of the instability regions along $\omega_0$ gets thinner and the deviation from $|\lambda_n|=1$ correspondingly weaker. Still we find that the instabilities persist even for a very high ratio $\omega_0/w$ in the conservative case considered here. However, once losses are included in the model, stability improves \cite{Arnold}. This explains why in the large $\omega_0$ calculations shown in the previous Sections instabilities were only observed at very small values of the magnetic flux.

Finally, in Panel~\textbf{B}, we show how the detailed structure of the Floquet quasi-energy spectra changes as the ratio $\omega_0/w$ is varied. To allow a direct comparison of the spectra, we have shifted each of these by the quantity $\bar{\omega}$ as defined in Eq.~\eqref{shift}. 
For large values of $\omega_0/w$ the spectra consists of $5$ bands in agreement with the HH model. 
As $\omega_0/w$ is decreased, the band structure of the effective Floquet Hamiltonian remains initially qualitatively similar to the one of the HH Hamiltonian with minor deviations due to the counter-rotating-wave terms. As $\omega_0/w$ is reduced further, the deviations become more significant and the band-gaps eventually close.  We have numerically checked that the topological edge states are no longer present once the bands-gaps have closed around $\omega_0/w\approx 7$.
   
\section{Experimental remarks}
\label{sec:experimental}

We conclude the work by commenting on possible experimental realizations of our proposal. As already mentioned, even though our attention was focussed on a classical mechanical implementation based on coupled pendula, all our results hold for harmonic oscillators of any type.

In all cases, the main challenge is to have a sufficiently high quality factor $Q=\omega_0/\gamma$. In fact, a high Q is crucial to have enough freedom to choose the frequencies of the system to satisfy well the following inequality chain:
\begin{equation}
\omega_0\gg s w \gg \Omega_{x,y} \gg \gamma.
\label{inequality2}
\end{equation}
These inequalities are required to be in the Rotating Wave limit, to find simple forms for the effective coupling amplitudes between neighbouring pendula, and to have clear access to the different spectroscopic features. 

As discussed around Fig.~\ref{fig:system_x}, the ratio $\omega_0/w$ also controls the number of sites $s$ that can be accommodated in a period of the spatial modulation. 
In particular, the sawtooth profile is needed for the scalability of the system of classical oscillators with a fixed quality factor $Q$. This can be seen by instead considering a simple linear ramp of the bare frequency of the pendula for a system of length $N_x$. In this case, the frequency difference between pendula would range up to $N_x w$ which, for a sufficiently large system, could exceed $\omega_0$ and thus violate Eq.~\eqref{inequality2}. 
The sawtooth modulation with finite $s$ is instead scalable to any arbitrarily large system while keeping the bare frequency $\omega_0$ to satisfy Eq.~\eqref{inequality2}.
We note that the atomic systems of~\cite{MonikaHH,MiyakeHH} can use a linear ramp of an external potential because they do not suffer from these difficulties, since counter-rotating-wave terms are not present. For optical cavities, instead, the bare cavity frequency and the $Q$ factor are typically large enough for this constraint to be practically irrelevant for realistic system sizes~\cite{modulatedCircuit}. 

While the spatial modulation of the oscillation frequency is rather straightforward to implement in pendula, a practical implementation of the temporal modulation needed for the Floquet scheme may be obtained, as sketched in~\cite{Salerno}, by mounting a magnet on each pendulum and letting it interact with an externally-controlled time-dependent magnetic field.

Quantitatively we need a minimum $Q$-factor of around $\omega_0/\gamma \approx 500$ 
to observe chiral edge states in the scheme that we propose, before encountering the band-closing of Fig.~\ref{fig:stability}-\textbf{B}. 
While reaching such a value of the Q factor in coupled pendula systems requires a significant, though still not dramatic, improvement of recent experimental set-ups~\cite{Huber}, it should be much more achievable using conventional RLC circuits. To go even further, the spectra of the HH model in Fig.~\ref{fig:spectra} and the Hofstadter butterfly in Fig.~\ref{fig:butterfly} were obtained using higher Q factors of $Q\approx 4000$ and $Q\approx 2\times 10^5$, respectively. Such high quality factors can be achieved in mechanical systems using micro-resonators as done in optomechanical oscillators~\cite{Marquardt}.
In fact, recently there has been a related proposal of realizing topological models using such optomechanical systems in the limit of the rotating-wave-approximation~\cite{Schmidt}.

\section{Conclusions and perspectives}
\label{sec:conclusions}

In this work, we have theoretically proposed a scheme to realize non-trivial topological models in classical systems of coupled pendula or coupled lumped-element RLC circuits through  a periodic time-modulation of system parameters. Within a Floquet framework, we have shown that suitably-designed temporal modulations of the natural oscillation frequencies effectively produce the complex coupling amplitudes of the HH model, where the synthetic magnetic flux is easily tunable.
We have assessed the accuracy of the Floquet picture by numerically solving the full equations of motion and we have shown that hallmarks of the HH model, such as the topologically-robust chiral edge states and the Hofstadter butterfly pattern, can be observed in the excitations of the system. Furthermore, we have illustrated a protocol to extract the topological Chern number of the bands in a classical counterpart of the integer quantum Hall effect. Finally, we extended our investigation to the regime where the analogy with the quantum lattice model breaks down, investigating the crucial role played by the counter-rotating-wave terms in the stability of the system and the persistence of topological effects. 

The exploration of topological effect in classical-mechanical systems is opening up many interesting new avenues of research. The use of Floquet modulation schemes to engineer topological models brings with it particular advantages. Firstly, time-modulation can naturally be used to break time-reversal symmetry and so to create edge states that are fully topologically protected. Secondly, the scheme discussed here is only one of many possible Floquet modulation schemes that could be applied to arrays of classical oscillators; other schemes may open up the way to build classical simulators of topological models with other interesting properties and with even more exotic topological invariants~\cite{Rudner,Rudner2}. 

Looking to the future, our proposal can be naturally extended to include a confining external potential, leading to novel oscillation features associated with analogue momentum-space magnetism~\cite{HannahPRL,TomokiHH,Andrei}. Beyond the small oscillation regime considered so far, systems of pendula also show nonlinear effects such as amplitude-dependent oscillation frequencies. The interplay of such nonlinearities with topological features may lead to a variety of new and unexpected effects.

\acknowledgments{We are grateful to N. Goldman, G. Ferrari, A. Berardo and N. Pugno for valuable discussions. This work was supported by the ERC through the QGBE grant, by the EU-FET Proactive grant AQuS, Project No. 640800, and by the Autonomous Province of Trento, partially through the project SiQuro. H.M.P was also supported by the EC through the H2020 Marie Sklodowska-Curie Action, Individual Fellowship Grant No: 656093 ``SynOptic".}

\appendix
\section{Full expressions of the effective equation of motion}
\label{app:fulleff}

In this appendix, we give more details on the derivation of the Hofstadter-like effective equation of motion in Eq.~\eqref{betadot}. 

From the Magnus expansion applied to time-periodic Hamiltonians at first order \cite{Bukov, GoldmanX, Goldman}, one gets the effective Hamiltonian as:
 $$\ham^\text{EFF}=\frac{1}{T}\int_0^T \ham(t) \mathrm{d}t+\dots$$ 
 for high modulation frequency. The effective Heisenberg equations that are derived from this effective Hamiltonian are $\dot{\beta}_{i,j}=-\ii [\beta_{i,j},\ham^\text{EFF}]=-\ii \frac{1}{T}\int_0^T [\beta_{i,j},\ham(t)]$. Therefore, from the equations of motion Eq.~\eqref{alphaRWAdot}, substituting the transformed variables of Eq.~\eqref{beta} and integrating over a period of the temporal modulation, the following effective equations of motion are found:

\begin{equation}
\begin{split}
\dot{\beta}_{i,j}&=-\ii \Delta \omega \beta_{i,j}  -\gamma_{i,j} \beta_{i,j} + \ii f_{i_p,j_p}^\text{EFF} \\& \quad +\ii \sum_{\pm1} \Omega_x^\text{EFF} \beta_{i\pm1,j}+\ii \sum_{\pm1} \Omega_y^\text{EFF} \beta_{i,j\pm1}.
\end{split}
\label{effdotbeta}
\end{equation}

The couplings along the $x$ and $y$ direction get renormalised, as does the external driving force. The expression for the effective coupling along $y$ is:
\begin{equation}
\Omega_y^\text{EFF}\equiv \Omega_y \sum_{p=-\infty}^{\infty} \mathcal{J}_{-(s-1)p}(I_0) \mathcal{J}_p(I_0) \e^{\ii \varphi_{i,j\pm1}ps}\e^{\ii (s-2p)\pi/2}
\label{effomegay}
\end{equation}
where we have used the Anger-Jacobi expansion to obtain the Bessel functions $\mathcal{J}_p(x)$ of $p$-th order with argument $x$. $I_0$ and $\varphi_{i,j}$ are defined as in Eq.~\eqref{argument} and Eq.~\eqref{fase}. 

The effective coupling along $x$ depends on the position along the static spatial modulation $S(i)$. We first focus on the couplings between pendula that have a natural-frequency difference of $\pm w$, where the $\pm$ indicates that the hopping is calculated going towards the left ($+1$) or towards the right ($-1$). We have that:
\begin{equation}
\begin{split}
&\Omega_x^{\text{EFF}(\pm w)} \equiv \Omega_x \sum_{p=-\infty}^{\infty} \mathcal{J}_{\pm 1 - (s-1)p}(I_0) \mathcal{J}_p(I_0) \\&\qquad\e^{-\ii\varphi_{i\pm1,j}(\pm1-p-(s-1) p)} \e^{\ii (\pm1-(s-1)p+p)\pi/2}.
\end{split}
\label{effomegax1}
\end{equation}
This equation describes the hopping along the small steps of the static spatial modulation. The coupling between pendula that have a natural-frequency difference of $\mp w(s-1)$ is:
\begin{equation}
\begin{split}
&\Omega_x^{\text{EFF}(\mp w(s-1))} \equiv \Omega_x \sum_{p=-\infty}^{\infty} \mathcal{J}_{(s-1)(\mp1-p)}(I_0) \mathcal{J}_p(I_0) \\&\qquad\e^{-\ii\varphi_{i\pm1,j}(\mp (s-1)-p-(s-1)p)} \e^{\ii (\mp(s-1)+p-(s-1)p)\pi/2}
\end{split}
\label{effeomegaxs}
\end{equation}
where this time $\mp$ indicates the hopping calculated going towards the left ($-1$) or towards the right ($+1$). This is because the frequency-difference along the ``big step'' of the static modulation has an opposite sign to the ones along the ``small steps''.
Finally, the effective driving force:
\begin{equation}
\begin{split}
f_{i_p,j_p}^\text{EFF} &\equiv f_{i_p,j_p}^\text{ex} \sum_{p'=-\infty}^{\infty} \mathcal{J}_{-(s-1){p'}}\left(\frac{V}{w}\right) \mathcal{J}_{p'}\left(\frac{V}{w}\right) \\&\quad \e^{-\ii(p'+(s-1)p')\phi_{i,j}}.
\end{split}
\label{feff}
\end{equation}

We now assume that $\mathcal{J}_0(I_0)  \gtrsim \mathcal{J}_1(I_0) \gg \mathcal{J}_p(I_0)$, with $p \geq 2$ 
and take only the largest term in the sums. In Eq.~\eqref{effomegay}, \eqref{effomegax1}, and \eqref{feff} we have that $p=p'=0$, while in Eq.~\eqref{effeomegaxs} we must take $p=\mp1$. Remarkably, in this approximation, the hopping along $x$ is uniform and does not depend on the position along the static-spatial modulation, thus allowing for clear definitions of $\Omega_x^\text{EFF}$, $\Omega_y^\text{EFF}$ and $f_{i_p,j_p}^\text{EFF}$:
\begin{equation}
\begin{split}
&\Omega_x^\text{EFF}=\Omega_y \,\mathcal{J}_{\pm1}\left(I_0\right) \mathcal{J}_0\left(I_0\right) \e^{\mp \ii (\varphi_{i\pm1,j}-\pi/2)}\\
&\Omega_y^\text{EFF}=\Omega_x \,\mathcal{J}_0\left(I_0\right)^2\\
&f_{i_p,j_p}^\text{EFF}=f_{i_p,j_p}^\text{ex} \,\mathcal{J}_0\left(V/w\right)^2.
\end{split}
\label{firstorder}
\end{equation}

By combining Eq.~\eqref{firstorder} with the effective equations of motion in Eq.~\eqref{effdotbeta} to lowest order in $I_0$ and $V$, we obtain exactly the equations in Eq.~\eqref{betadot}. 

From Eq.~\eqref{effomegay},\eqref{effomegax1} and \eqref{effeomegaxs} we notice that the second largest term in the sums is proportional to $\mathcal{J}_{(s-1)}(I_0)$, that is very small if $I_0 \ll 1$ or when considering a large period of the static modulation $s\gg 2$. We found that an optimal combination of these two requirements that fulfils the inequality Eq.~\eqref{inequality} with the renormalisation of Eq.~\eqref{renormalisedOmega} is to have $I_0=0.5$ with $s=5$.

\section{Fourier decomposition of the equations of motion}
\label{app:Fourier}

We now comment on the Fourier decomposition that was used as an alternative method to the full-time integration with Runge-Kutta for solving the set of differential equations in Eq.~\eqref{alphadrivendot} to produce the butterfly spectra in Fig.~\ref{fig:butterfly}. 

From the properties of periodically driven systems \cite{GoldmanX,Goldman}, we know that, during a Floquet evolution between two stroboscopic times $t_n=nT$, the system exhibits micro-motion, with a period set by the frequency of the temporal modulation $w$. We are interested in the steady state, therefore, we search for a solution of Eq.~\eqref{alphadrivendot} that oscillates at the frequency of the external driving $\omega_\text{ex}$. To this end, we expand the $\alpha_{i,j}(t)$ in a Fourier series, using both $\omega_\text{ex}$ and $w$ as harmonics:
\begin{equation}
\alpha_{i,j}(t)=\sum_{m=-\infty}^\infty \sum_{n=-\infty}^\infty \alpha_{i,j}^{(m,n)} \e^{\ii m w t} \e^{\ii n\omega_\text{ex}t}
\label{Fourier}
\end{equation}
where $\alpha_{i,j}^{(m,n)}$ are the time-independent Fourier amplitudes, assuming that all the time-dependencies are in the exponential term. 
The sum over $m$ is truncated to a finite number $M$, that is large enough to ensure a convergent solution. The sum over $n$ takes only two values $\pm1$, since we have assumed that the external driving force is a cosine with a defined frequency $\omega_\text{ex}$. By substituting  Eq.~\eqref{Fourier} in Eq.~\eqref{alphadrivendot} and isolating the component proportional to $\e^{\ii m w t } \e^{\ii n\omega_\text{ex}t}$, we have a set of linear algebraic equations that can be inverted to find the coefficients $\alpha_{i,j}^{(m,n)}$. 

In order to simulate the real experimental situation, we performed all the calculations using the full numerical integral of Eq.~\eqref{alphadrivendot}, except for Hofstadter butterfly where the full numerical integration would have been computationally demanding. We have verified the good agreement between the two methods finding a mean error that is less than $1\%$. We note that the two methods are in fact equivalent when the system is dynamically stable. However, when there is an instability, the system does not reach the steady state. This is clearly seen in the full numerical integration method, where the solution shows the typical exponential growth of an unstable system. In the Fourier method the steady state is imposed by the decomposition Eq.~\eqref{Fourier} itself and so the emergence of an instability can not be predicted, as discussed in the main text.

\end{document}